\documentclass[12pt,a4paper]{article}
\pdfoutput=1

\usepackage[T1]{fontenc}

\usepackage{jheppub}
\usepackage{psfrag}
\usepackage{cancel}
\usepackage{lscape}
\usepackage{slashed}
\usepackage{mathtools}

\usepackage{amssymb}
\usepackage{amsmath}

\usepackage{caption}
\usepackage{array}
\usepackage{graphicx}
\usepackage{subcaption}
\usepackage{multirow}

\def\dd{d\!\!{}^-\!}
\def\del{\delta\!\!\!{}^-\!}

\usepackage[utf8]{inputenc}
\usepackage[textwidth=22mm]{todonotes}

\DeclareMathOperator{\tr}{tr}

\preprint{{Edinburgh 2017/25, QMUL-PH-17-25}}

\title{Inelastic Black Hole Scattering from Charged Scalar Amplitudes}

\author[a]{Andr\'es Luna,}
\author[b,c]{Isobel Nicholson,}
\author[b,c]{Donal O'Connell,}
\author[d]{Chris D. White}
\affiliation[a]{School of Physics and Astronomy, University of Glasgow, G12 8QQ, Scotland, UK}
\affiliation[b]{
Higgs Centre for Theoretical Physics, School of Physics and Astronomy, The University of Edinburgh, Edinburgh EH9 3JZ, Scotland, UK%
}
\affiliation[c]{Kavli Institute for Theoretical Physics, University of California, Santa Barbara, CA 93106-4030, USA}
\affiliation[d]{Centre for Research in String Theory, School of Physics and Astronomy, Queen Mary University of London, 327 Mile End Road, London E1 4NS, UK}
\emailAdd{a.luna-godoy.1@research.gla.ac.uk,i.nicholson@sms.ed.ac.uk,\\donal@ph.ed.ac.uk, christopher.white@qmul.ac.uk}

\abstract{
We explain how the lowest-order classical gravitational radiation produced during the inelastic scattering of two Schwarzschild black holes in General Relativity can be obtained from a tree scattering amplitude in gauge theory coupled to scalar fields. The gauge calculation is related to gravity through the double copy. We remove unwanted scalar forces which can occur in the double copy by introducing a massless scalar in the gauge theory, which is treated as a ghost in the link to gravity. We hope these methods are a step towards a direct application of the double copy at higher orders in classical perturbation theory, with the potential to greatly streamline gravity calculations for phenomenological applications.
}

\keywords{Scattering amplitudes in gauge theory and
  gravity, gravitational radiation}

\begin{document}
\maketitle
\flushbottom

\section{Introduction \label{sec:intro}}

General Relativity is a spectacularly successful description of gravitational processes. It is also celebrated for its great beauty. However, the perturbative expansion of the Einstein-Hilbert Lagrangian about Minkowski space is a less well-beloved aspect of Einstein's theory. Because of the presence of the metric and its inverse, as well as the square root of the determinant of the metric, the perturbative expansion contains an infinite number of terms, each of which corresponds to a complicated Feynman rule. This complexity makes calculation difficult.

Gradually we have realised that a surprising approach to gravitational perturbation theory about Minkowski space has the potential to greatly improve this situation. The basis of the idea goes back to the early days of string theory, when Kawai, Lewellen and Tye (KLT) realised~\cite{Kawai:1985xq} that closed string amplitudes can be obtained from open string amplitudes. In the field theory limit, this means that graviton scattering amplitudes may be obtained from knowledge of gauge boson scattering amplitudes, and of the KLT relations. 

One disadvantage of the KLT connection between gauge theory and gravity is that the KLT relations themselves are quite complicated, especially for processes involving many gravitons. They are also restricted to tree level scattering amplitudes, so that their applicability to loop processes is necessarily indirect. Fortunately, another perspective on the connection between gauge theory and gravity, known as the Bern, Carrasco and Johansson (BCJ) double copy, removes both of these disadvantages~\cite{Bern:2008qj,Bern:2010ue}. The double copy is based on a simple rearrangement of Yang-Mills amplitudes which has a distinctly group theoretic feel. It also has an immediate application to loop diagrams.

To date, the double copy has been used primarily to understand the quantum structure of gravity. But there is also considerably motivation to understand classical processes in General Relativity to high precision. This motivation comes from future plans for high precision gravitational wave observatories, such as LISA and the Einstein telescope. Does the double copy provide a route to a refined understanding of radiative processes involving black holes?
In fact, scattering amplitudes have already been used to extract information about the classical scattering processes in gravity. One direction is to determine the effective potential between objects deduced from two-to-two amplitudes, which has been studied in a series of remarkable papers which build on the full suite of amplitudes tools~\cite{Neill:2013wsa,Bjerrum-Bohr:2013bxa,Bjerrum-Bohr:2014lea,Bjerrum-Bohr:2014zsa,Bjerrum-Bohr:2016hpa,Cachazo:2017jef,Guevara:2017csg}, see also~\cite{Bjerrum-Bohr:2017dxw} for a recent pedagogical introduction. There is clearly great scope for further work in this area~\cite{Damour:2017zjx}.

However, the double copy is set up to compute graviton scattering
amplitudes. Perhaps it may be possible to directly compute the
classical gravitational wave spectrum in a scattering process directly
from the double copy? Indeed, the situation seems to be very
positive. The structure of the double-copy is reflected in a class of
solutions of the Einstein equations, known as Kerr-Schild
solutions. These simple, symmetric spacetimes are associated with
similarly simple and symmetrical exact solutions of the Yang-Mills
equations in a classical manifestation of the double
copy~\cite{Monteiro:2014cda,Luna:2015paa,Luna:2016due,Ridgway:2015fdl,Adamo:2017nia,Adamo:2017sze,Bahjat-Abbas:2017htu,Carrillo-Gonzalez:2017iyj}. Similarly,
perturbative spacetimes can be constructed order-by-order in a manner
which directly manifests the double copy~\cite{Luna:2016hge}. But it
was still a wonderful surprise to see genuine gravitational
scattering, with the production of gravitational radiation, emerge in
a remarkably simple form from the double copy as shown recently by
Goldberger and Ridgway~\cite{Goldberger:2016iau}, see
also~\cite{Goldberger:2017frp}.

One disadvantage of the Goldberger-Ridgway calculation is the presence
of unwanted fields in the classical theory. In the double copy it is a
simple fact that the graviton naturally comes along with two extra
fields: a scalar dilaton and an antisymmetric tensor known as the axion.  In any given calculation, we can
hope to switch the dilaton and axion off, but typically in the
simplest cases they will be present. Indeed, in the simplest
perturbative construction of a spacetime containing a point-like
singularity using the double copy~\cite{Luna:2016hge}, the dilaton
couples to the singularity with strength proportional to the
gravitational mass of the singularity. The spacetime is therefore a
JNW naked singularity~\cite{Janis:1968zz} rather than the
Schwarzschild solution. As a consequence, the objects scattering in
the Goldberger-Ridgway calculation were JNW singularities. 

Another
interesting aspect of reference~\cite{Goldberger:2017frp} was how the authors
implemented the double copy. Rather than following the standard
prescription from scattering amplitudes, they
replaced Yang-Mills colour factors with the kinematic part of the
Yang-Mills three point vertex. This is remarkable in view of the fact
that it is
well-known~\cite{Bern:2010yg,Du:2011js,BjerrumBohr:2012mg,Fu:2012uy,Bjerrum-Bohr:2013iza,Du:2013sha,Tolotti:2013caa,Nohle:2013bfa,Monteiro:2013rya,Fu:2014pya,Mastrolia:2015maa,Fu:2016plh,Brown:2016mrh}
that the four point vertex in gauge theory is important for
colour-kinematics to work in general.

The goal of this paper is to understand classical inelastic gravitational scattering
from the point of view of scattering amplitudes, given that it is in this
context that the double copy has its most
well-understood form. In particular, we will show how to obtain
gravitational radiation by copying a Yang-Mills amplitude. Working at lowest order,
it is sufficient to consider the
scattering of a pair of scalar particles, which represent non-spinning (Schwarzschild) black holes. We will show
explicitly how the Goldberger-Ridgway approach can be related to
scattering amplitudes, and comment on
a particular simplification which occurs in the double copy in this specific context.
Furthermore, the amplitude approach can be used to remove the
contribution of the dilaton. To this end, we build on the recent work
of Johansson and Ochirov~\cite{Johansson:2014zca}, who obtained pure
gravity as a double copy. 

We hope that our results represent a step toward
an application of the double copy in more detailed calculations of gravitational phenomena.
Evidently the double copy has the capacity to determine genuine perturbations to the metric 
about Minkowski space resulting from physical dynamical processes. The technical simplicity of
the double copy could
greatly streamline the calculation of phenomenologically relevant
quantities if it can be applied to physically relevant processes.

The structure of our paper is as follows. In section~\ref{sec:review}
we recall salient details regarding BCJ duality and the double copy,
as well as reviewing the classical scattering calculations of
refs.~\cite{Goldberger:2016iau}. In section~\ref{sec:gaugeScalarAmp}
we show how these results can be reproduced using scattering
amplitudes, which will involve a detailed discussion of taking
appropriate kinematic limits to make this equivalence manifest. In
section~\ref{sec:dilaton}, we demonstrate explicitly how the dilaton
can be removed to generate results in pure gravity, building on the
ideas of~\cite{Johansson:2014zca}. Finally, we discuss our results and
conclude in section~\ref{sec:conclusions}.

\section{Review \label{sec:review}}

To keep the article self-contained, we open with a review of the BCJ
story: colour-kinematics duality and the double copy. We have also
included a brief review of reference~\cite{Goldberger:2017frp} which will
be important for the remainder of the paper.

\subsection{Review of BCJ}

The essence of BCJ is an understanding of how gauge invariance works in the scattering amplitudes of Yang-Mills theory. We can always choose to write these amplitudes as a sum over Feynman-like diagrams with three-point vertices. If the set $\Gamma$ contains all of these diagrams with $n$ external points, then the $n$-point amplitude is
\begin{equation}
\mathcal{A} = \sum_{i \in \Gamma} \frac{c_i n_i}{d_i},
\label{eq:generalYMAmp}
\end{equation}
where $d_i$ is the Feynman propagator denominator associated with graph $i$, while $c_i$ is the Yang-Mills colour factor corresponding to the diagram and $n_i$ is the kinematic numerator of the graph. Notice that only the kinematic numerator $n_i$ depends on the polarisations of the particles. Let us choose one particle, say particle 1, and replace its polarisation vector $\epsilon_1$ by its momentum $p_1$. In that case gauge invariance requires that the amplitude $\mathcal{A} = 0$. However, it is not the case that all the numerators $n_i = 0$. Rather, the identity $\mathcal{A} = 0$ follows from a cancellation among the distinct diagrams. This is only possible because the colour factors $c_i$ are not all independent. Instead, they obey Jacobi identities which arise from pure group theory.

It is also the case that the $n_i$ are not all uniquely defined. Indeed, there is a large space of numerators which have the property that~\eqref{eq:generalYMAmp} is a valid expression for the Yang-Mills amplitude. The idea of BCJ is then to pick very special numerators which have the property that whenever graphs $\alpha$, $\beta$ and $\gamma$ are such that the colour factors satisfy a Jacobi identity,
\begin{equation}
c_\alpha \pm c_\beta \pm c_\gamma = 0,
\end{equation}
then the kinematic numerators satisfy the same identity,
\begin{equation}
n_\alpha \pm n_\beta \pm n_\gamma = 0.
\end{equation}
Notice that we have allowed for the possibility that there may be positive and negative signs in the Jacobi identity; whatever these signs are, they must be in common between the colour and kinematic identities. We also require that the kinematic numerators satisfy the same antisymmetries as the colour factors. These requirements are known as colour-kinematics duality, because both colour and kinematics have the same algebraic properties.

The reason for making this choice is that we may now construct a new amplitude which must be gauge invariant and local:
\begin{equation}
\mathcal{M} = \sum_{i \in \Gamma} \frac{n_i n_i}{d_i}.
\label{eq:symmetricGravAmp}
\end{equation}
Comparing to our previous gauge amplitude, we have replaced the colour factors $c_i$ with a second copy of the kinematic numerators, $n_i$. For this reason, equation~\eqref{eq:symmetricGravAmp} is known as the double copy formula. The quantity $\mathcal{M}$ is gauge invariant because, if we replace the polarisation vector $\epsilon_1$ with $p_1$ in, say, the left $n_i$ factors, then $\mathcal{M} = 0$. The identity must follow from precisely the same algebra as in gauge invariance of the Yang-Mills amplitude. Similarly we could replace the polarisation vector with the momentum in just the right $n_i$ factor. Locality is assured because we have included precisely the correct Feynman denominators for a local field theory. Therefore $\mathcal{M}$ is a scattering amplitude of some kind. 

It is straightforward to see that $\mathcal{M}$ is an amplitude in a theory of gravity. To see this, notice that for each particle, $\mathcal{M}$ is linear in the outer product $\epsilon_i^\mu \epsilon_i^\nu$ of polarisation vectors. We may decompose this outer product into irreducible representations of the little group. The symmetric, traceless tensor is the polarisation tensor of a gravitational wave. Other kinds of particles are present: the trace term in the tensor product decomposition corresponds to a massless scalar particle which is known as the dilaton, while the antisymmetric tensor is known as the axion. The presence of these states is a natural feature of the double copy. 
Since in the double copy, the graviton, the axion and the dilaton all emerge from
a tensor product decomposition of one matrix, it is useful to give
them a collective name: we will refer to them as the product
graviton\footnote{The term {\it fat graviton} was used in
  ref.~\cite{Luna:2016hge} for this quantity. Here we wish to avoid
  ambiguity due to a similar term being used elsewhere in the
  literature.}. 

More generally, we can allow for two different choices of $n_i$ in the double copy, for example numerators of different gauge theories:
\begin{equation}
\mathcal{M} = \sum_{i \in \Gamma} \frac{n_i \tilde n_i}{d_i}.
\label{eq:generalGravAmp}
\end{equation}
In this way we can construct amplitudes corresponding to a variety of theories. There is an active programme of research aimed at determining what kinds of gravitational theories can be constructed from the double copy~\cite{Chiodaroli:2017ehv,Anastasiou:2017nsz,Chiodaroli:2017ngp,Anastasiou:2016csv,Chiodaroli:2015wal,Chiodaroli:2015rdg,Chiodaroli:2014xia,Chiodaroli:2013upa,Carrasco:2012ca}. The construction of pure Einstein gravity as a double copy, due to Johansson and Ochirov~\cite{Johansson:2014zca} is a particularly interesting case, and it is one which will play a central role in the present paper. 

It is also worth emphasising that the double copy has an immediate
extension to loop amplitudes~\cite{Bern:2010ue}. This fact has led to
a wealth of progress in our understanding of the structure of
(super)gravity~\cite{Johansson:2017bfl,Bern:2017tuc,Bern:2017puu,Yang:2016ear,Bern:2014lha,Bern:2014sna,Bern:2013uka,Bern:2013qca,Bern:2013yya,Bern:2012gh,Bern:2012cd,Bern:2012uf,Bern:2011rj,Bern:2011ia},
most recently including the integrand of the five-loop, four-point
amplitude in maximal supergravity in four
dimensions~\cite{Bern:2017ucb}. This last achievement rested on a new
general technique~\cite{Bern:2017yxu} for constructing appropriate
numerators which we anticipate will be very useful in the
future. All-order evidence for the double copy has been obtained in
special kinematic
limits~\cite{Oxburgh:2012zr,Melville:2013qca,Luna:2016idw,Saotome:2012vy,Johansson:2013nsa,Vera:2012ds}.

The double copy rests on colour-kinematics duality, which hints at the existence of a new kind of symmetry in gravity---a kinematic symmetry which controls the structure of the kinematic numerators. This kinematic algebra remains mostly mysterious, except in the context of the self-dual theory~\cite{Monteiro:2011pc} and the nonlinear sigma model~\cite{Cheung:2016prv}. But in spite of the slow progress in our understanding of the kinematic algebra, the double copy provides a new way of understanding the symmetry structure of supergravity as following from symmetries of Yang-Mills theory~\cite{Cardoso:2016amd,Cardoso:2016ngt,Anastasiou:2015vba,Nagy:2014jza,Anastasiou:2014qba,Anastasiou:2013hba,Borsten:2013bp}.

\subsection{Classical gravitational scattering}
\label{sec:goldberger}
In this section, we review the work of Goldberger and Ridgway~\cite{Goldberger:2016iau}, who computed the gravitational radiation emitted during the inelastic scattering of two JNW singularities using a method based on the double-copy. Thus, their computation began in the context of gauge theory. Specifically, they considered classical, coloured point particles with positions $x_i(\tau)$ and colours $c_i^a(\tau)$ and masses $m_i$ interacting through a gauge field $A^a_\mu$. If we denote the coupling by $g$ and let $F_{\mu\nu}$ be the gauge field strength tensor, the classical equations defining the system are
\begin{subequations}
\begin{align}
D^\mu F^a_{\mu\nu} &= g \sum_i \int d \tau c_i^a(\tau) \frac{dx_i^\nu(\tau)}{d\tau} \delta^{(d)}(x - x_i(\tau)), \\
m_i \frac{d^2 x_i^\mu(\tau)}{d \tau^2} &= g F^{a\mu\nu} c^a_i(\tau) \frac{dx_{i\nu}(\tau)}{d\tau}, \\
\frac{dc_i^a(\tau)}{d\tau} &= g f^{abc} \, \frac{dx_i^\mu}{d\tau} A_\mu^b(x_i(\tau)) \, c^c_i(\tau).
\end{align}
\label{eq:classicalEqns}
\end{subequations}

When the scattering angle is small, one can solve these equations order-by-order in perturbation theory. At zeroth order, the particles move on straight-line trajectories, with constant velocity $v_i^\mu$ and constant colour:
\begin{align}
x_i^\mu(\tau) &\simeq b_i^\mu + v_i^\mu \tau, \\
c_i^a(\tau) & \simeq c_i^{(0)a}.
\end{align}
Interactions correct these expressions, leading to perturbative expansions of the positions, the colours, and indeed of the field. We will indicate terms arising at the $n$th order of perturbation theory with a superscript in brackets:
\begin{align}
x_i^\mu(\tau) &= x_i^{(0)\mu}(\tau) + x_i^{(1)\mu}(\tau) + \cdots, \\
c_i^a(\tau) &= c_i^{(0)a} + c_i^{(1)a}(\tau) + \cdots, \\
A_\mu^a(x) &= A_\mu^{(0)a}(x) +  A_\mu^{(1)a}(x) + \cdots.
\end{align}

With this setup, it is a mechanical task to perturbatively solve the equations of motion to any desired accuracy. We are interested in radiation emitted during a collision, which requires us to compute $A_\mu^{(1)a}(x)$. The result is
\begin{multline}
k^2 A^{(1)a\mu}(k) = g^3  \int \dd q_1 \dd q_2 \del(k - q_1 - q_2) \del(q_1 \cdot v_1) e^{i q_1 \cdot b_1}  \del(q_2 \cdot v_2) e^{i q_2 \cdot b_2}  \\
\times\left\{\frac{c_1^{(0)a}}{m_1} \frac{c_1^{(0)} \cdot c_2^{(0)}}{k\cdot v_1 \; q_2^2} \left[ -v_1 \cdot v_2 \left(q_2^\mu - \frac{k \cdot q_2}{k \cdot v_1} v_1^\mu\right)  + k \cdot v_1 \; v_2^\mu -k \cdot v_2 \; v_1^\mu \right] \right. \\
\left. + \frac{i f^{abc} c_1^b c_2^c}{q_1^2 q_2^2} \left[ 2 k\cdot v_2 \; v_1^\mu -v_1 \cdot v_2 \; q_1^\mu + v_1 \cdot v_2 \frac{q_1^2}{k\cdot v_1} v_1^\mu\right] + \left( 1 \leftrightarrow 2 \right) \right\}.
\label{eq:classicalA}
\end{multline}
Of course, our main focus is not  gauge radiation but rather the gravitational radiation obtained via the double copy. Goldberger and Ridgway implemented the double copy as a set of replacement rules:
\begin{align}
\label{eg:grReplacement}
c_i^{(0)a} &\rightarrow m_i v_i^{\mu} ,\\
if^{abc} &\rightarrow \frac12 \Gamma^{\mu\nu\rho}(q_1, q_2, q_3 ).
\label{eg:grReplacement2}
\end{align}
In the latter replacement, the quantity $\Gamma$ is proportional to the Yang-Mills three point amplitude, while the three momenta $q_1$, $q_2$ and $q_3$ are momenta associated with the lines with colours $a$, $b$ and $c$. Specialising to the two particle case, the result is a perturbative ``product'' graviton given by
\begin{multline}
k^2 H^{(1)\mu\nu}(k) = - \frac{m_1 m_2}{8 m_\textrm{pl}^{3(d-2)/2}} \int \dd q_1 \dd q_2 \del(k - q_1 - q_2) \del(q_1 \cdot v_1) e^{i q_1 \cdot b_1}  \del(q_2 \cdot v_2) e^{i q_2 \cdot b_2} \\\left[ \frac{v_1 \cdot v_2}{q_2^2 \; k \cdot v_1} v_1^\nu \left\{ v_1 \cdot v_2 \left( \frac12 (q_2 - q_1)^\mu - \frac{k \cdot q_2}{k \cdot v_1} v_1^\mu \right) + k \cdot v_2 \; v_1^\mu - k \cdot v_1 \; v_2^\mu \right\} \right. \\ \left. + \frac{2 k \cdot v_2 \; v_1^\nu - 2 k \cdot v_1 \; v_2^\nu + v_1 \cdot v_2 (q_2 - q_1)^\nu }{2 q_1^2 q_2^2}  \left( 2 k \cdot v_2 \; v_1^\mu - v_1 \cdot v_2 \; q_1^\mu + \frac{v_1 \cdot v_2 \; q_1^2}{k \cdot v_1} v_1^\mu \right) \right. \\ \left. \vphantom{\frac{v_1 \cdot v_2}{k \cdot v_1} }+ (1 \leftrightarrow 2) \right].
\end{multline}
It was demonstrated in reference~\cite{Goldberger:2017frp}, by direct calculation, that this
product graviton encodes the gravitational radiation emitted in the
scattering of two JNW singularities.

For our purposes, it is helpful to exploit the symmetry in particles 1 and 2 to slightly rewrite $H^{(1)\mu\nu}(k)$ in a manner which makes gauge invariance more manifest. To that end, we introduce the vectors
\begin{subequations}
\begin{align}
P_{12}^\mu &\equiv k \cdot v_1 \; v_2^\mu - k \cdot v_2 \; v_1^\mu, \\
Q_{12}^\mu &\equiv (q_1-q_2)^\mu- \frac{q_1^2}{k \cdot v_1} v_1^\mu+ \frac{q_2^2}{k \cdot v_2} v_2^\mu,
\end{align}
\label{eq:PQ}
\end{subequations}
which are gauge invariant in the sense that $P_{12} \cdot k = 0 = Q_{12} \cdot k$. The product graviton can be written as
\begin{multline}
\label{eq:classical}
k^2 H^{(1)\mu\nu}(k) = - \frac{m_1 m_2}{8 m_\textrm{pl}^{3(d-2)/2}} \int \dd q_1 \dd q_2 \del(k - q_1 - q_2) \del(q_1 \cdot v_1) e^{i q_1 \cdot b_1}  \del(q_2 \cdot v_2) e^{i q_2 \cdot b_2} \times \\
\left[
\frac{P_{12}^\mu P_{12}^\nu}{q_1^2 q_2^2} +\frac{v_1 \cdot v_2}{2 q_1^2 q_2^2}  \left( Q_{12}^\mu P_{12}^\nu + Q_{12}^\nu P_{12}^\mu\right) 
+\frac{(v_1 \cdot v_2)^2}4  \left(\frac{Q_{12}^\mu Q_{12}^\nu}{q_1^2 q_2^2 } - \frac{P_{12}^\mu P_{12}^\nu}{(k \cdot v_1)^2 (k\cdot v_2)^2} \right) 
\right].
\end{multline}
A number of questions arise from this calculation, such as:
\begin{enumerate}
\item Are the double-copy replacement rules in equations~\eqref{eg:grReplacement} the same as the BCJ rules, or a replacement for them? What about colour-kinematics duality?
\item Can we find a straightforward mechanism for removing the dilaton pollution in the calculation?
\end{enumerate}
To address these questions, we find it convenient to reformulate the black hole scattering calculation as a scattering amplitude. 

\section{Charged Scalar Amplitudes\label{sec:gaugeScalarAmp}}

\begin{figure}[t]
\centering
	\includegraphics[height=0.2\textheight]{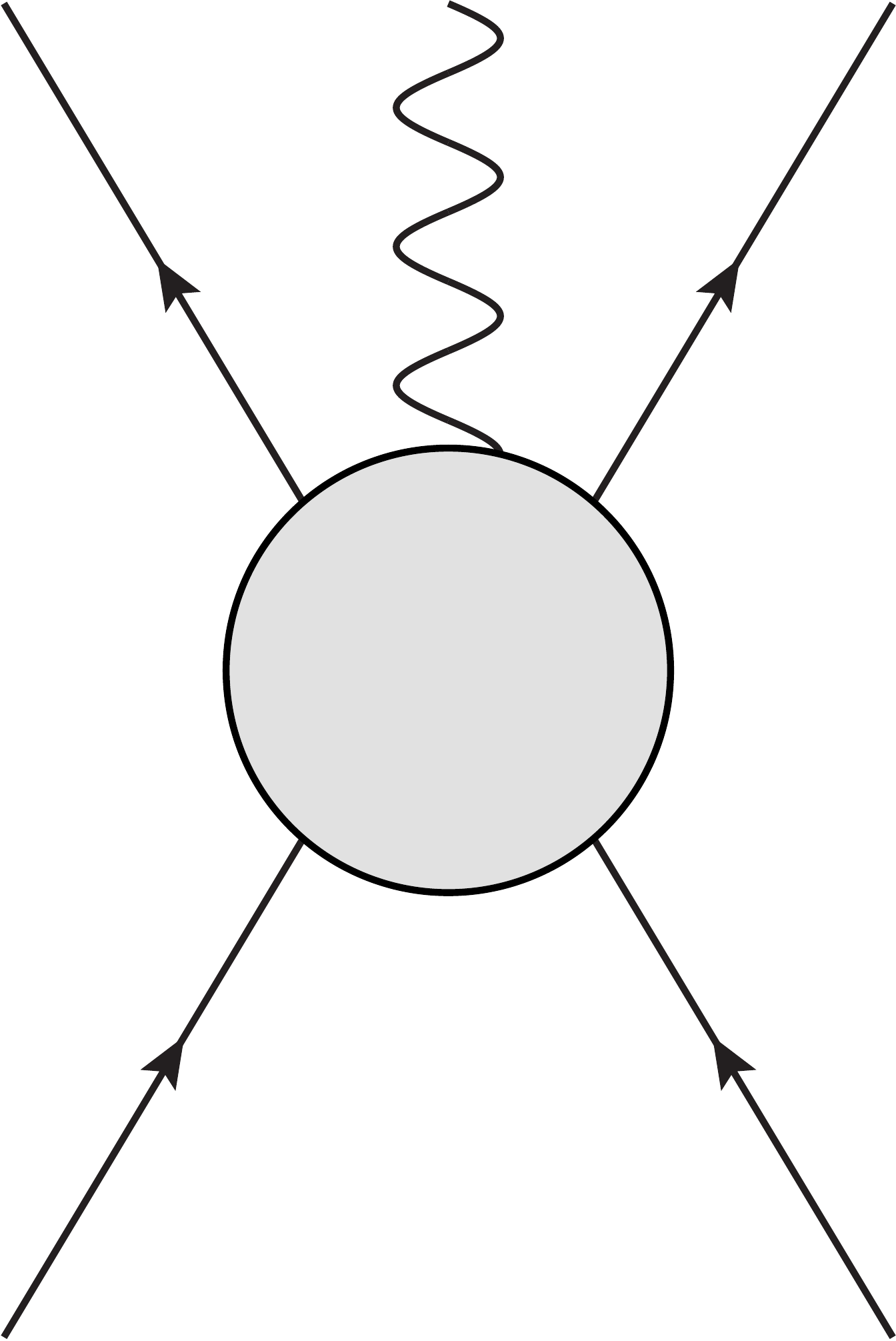}
	\caption{Two particle scattering with the production of radiation. This is the basic structure of diagrams we will be interested in. Note that time is on the vertical axis.}
	\label{fig:blob}
\end{figure}

Our aim is to recast the emission of gravitational radiation from a
pair of scattering particles, in terms of a scattering amplitude
calculation. Amplitudes have the advantage that the application of the
double copy is well-established, as is the possibility of removing
unwanted dilaton contributions. To this end, we must first decide what
scattering amplitude to calculate. We will begin in Yang-Mills theory,
given that we wish to obtain the gravity result using the double
copy. The simplest possible candidate is then a five-point amplitude,
corresponding to the incoming / outgoing particles, plus an additional
gluon, as shown in figure~\ref{fig:blob}. The scattering particles
themselves, however, need not be gluons. Ultimately, our gravity
calculation will describe the scattering of astrophysical objects
(e.g. black holes) of arbitrary spin. Thus, we must add additional
matter to our pure Yang-Mills theory, whose spin is directly related
to the spin of the objects whose scattering we wish to study. Given
that our main motivation is to illustrate the double copy and removal
of the dilaton, we will restrict ourselves to scalar scattering
particles in what follows. 

The masses of our two incoming particles need not be the same, so we
will consider a gauge field $A^a_\mu$ coupled to two different massive
scalars $\Phi_i$ transforming in representations $R_i$ of the gauge
group. The Lagrangian is simply
\begin{equation}
\mathcal{L} = -\frac12 \tr F^{\mu\nu} F_{\mu\nu} + \sum_i \left[(D_\mu \Phi_i)^\dagger (D^\mu \Phi_i) - m^2_i |\Phi|^2\right].
\end{equation}
A discussion of how the double copy works in this kind of situation can be found in~\cite{Johansson:2014zca}. The double copy of our five point gauge amplitude will be an amplitude in a theory containing gravity, a dilaton, an axion, and the scalar fields\footnote{If any of the representations $R_i$, say $R_1$ were to be the adjoint, it may be appropriate to include vector states in the double copy built up from one gauge field times one $\Phi_1$. However in this case we may choose the representations at will and avoid these unwanted states. We thank Radu Roiban for discussions on this point.}.

The goal is to reproduce a lowest order calculation in classical field
theory. Since we aim for a classical result, you might think that it
is obvious that we should only compute tree diagrams: after all, it is
standard lore that loops are quantum corrections! However, this is not
quite accurate. We will indeed compute a tree diagram, but the
justification is that we wish for a \emph{lowest} order result in
classical field theory. Loops are relevant for higher orders in the
classical limit when massive particles are
present~\cite{Holstein:2004dn}.

\subsection{The scattering amplitude \label{sec:fullAmp}}

\begin{figure}
\centering
	\begin{subfigure}[t]{0.32 \textwidth}
		\includegraphics[width=0.9\textwidth]{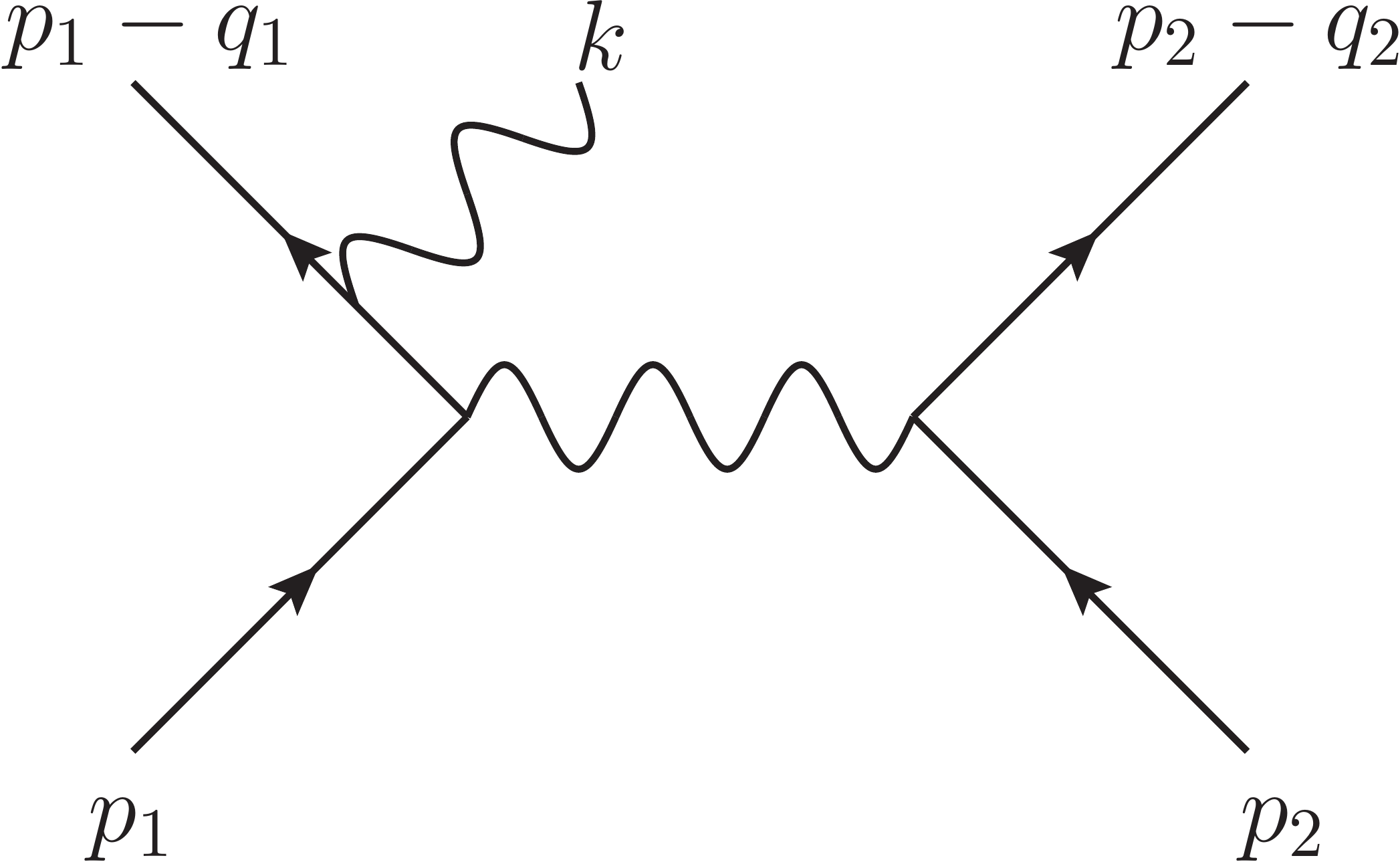}
		\caption{Diagram A}
		\label{fig:scalarDiagramsA}
	\end{subfigure}
	\begin{subfigure}[t]{0.32\textwidth}
		\includegraphics[width=0.9\textwidth]{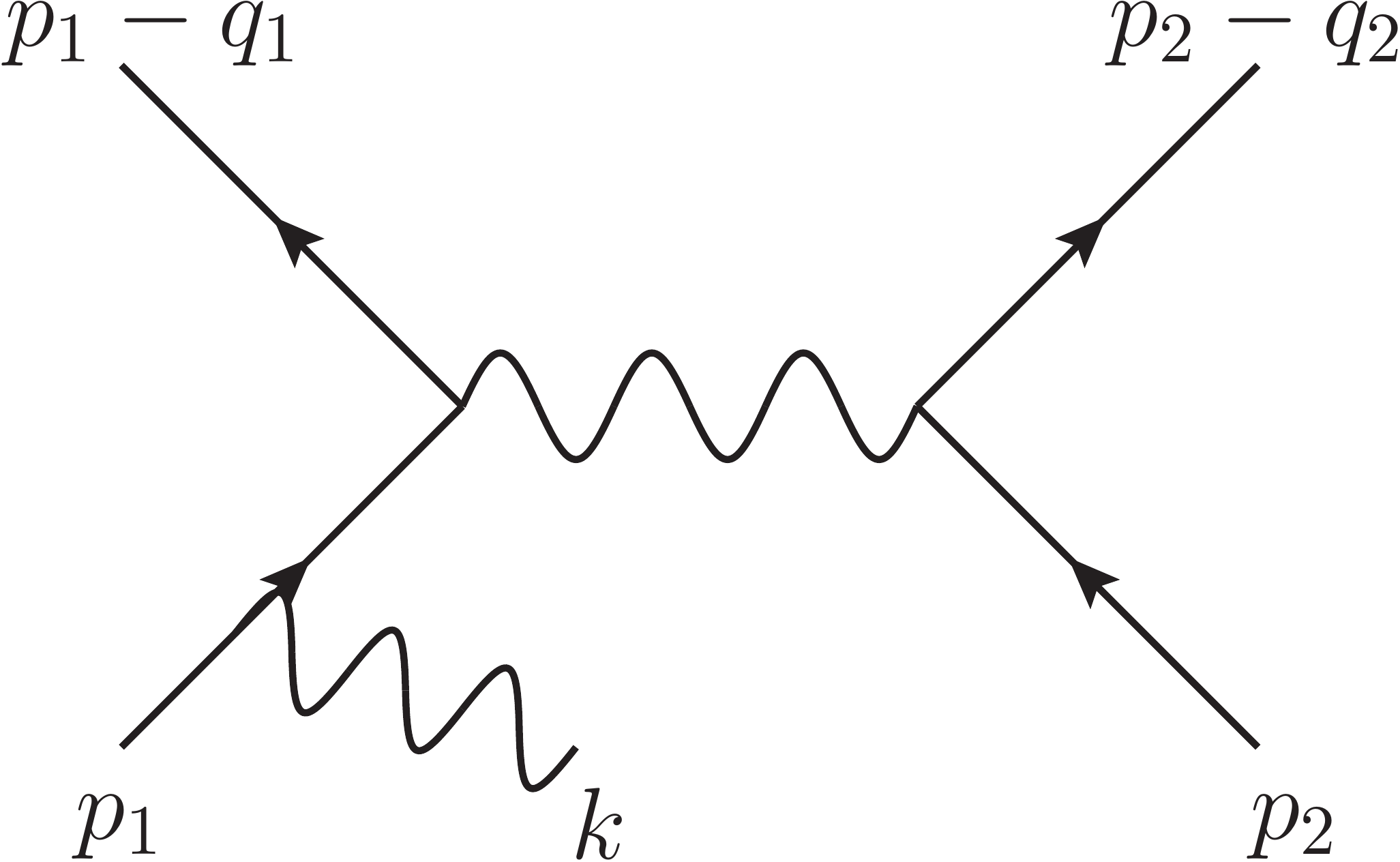}
		\caption{Diagram B} 
		\label{fig:scalarDiagramsB}
	\end{subfigure}
	\begin{subfigure}[t]{0.32\textwidth}
		\includegraphics[width=0.9\textwidth]{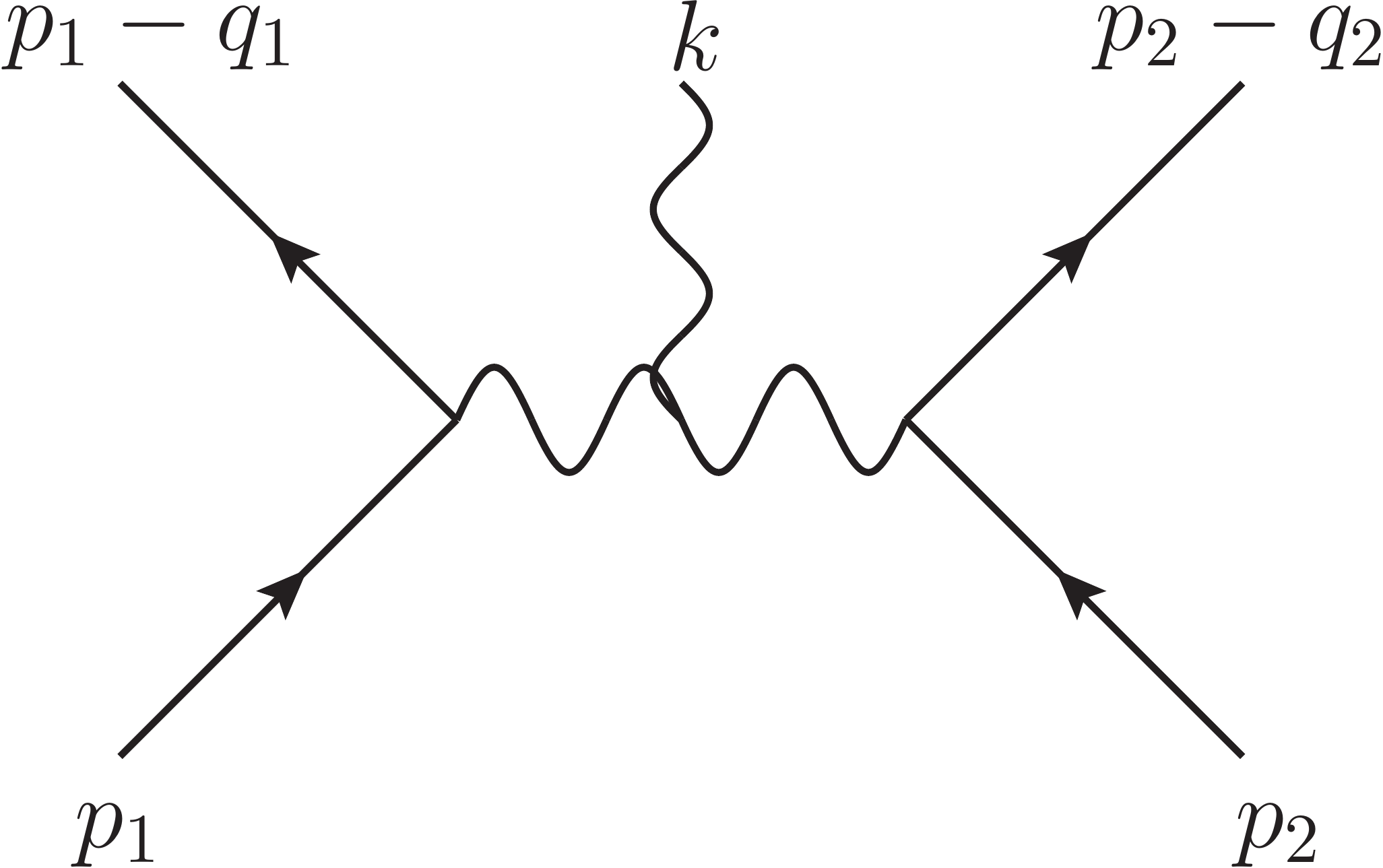}
		\caption{Diagram C}
		\label{fig:scalarDiagramsC}
	\end{subfigure}
	\begin{subfigure}[t]{0.32\textwidth}
		\includegraphics[width=0.9\textwidth]{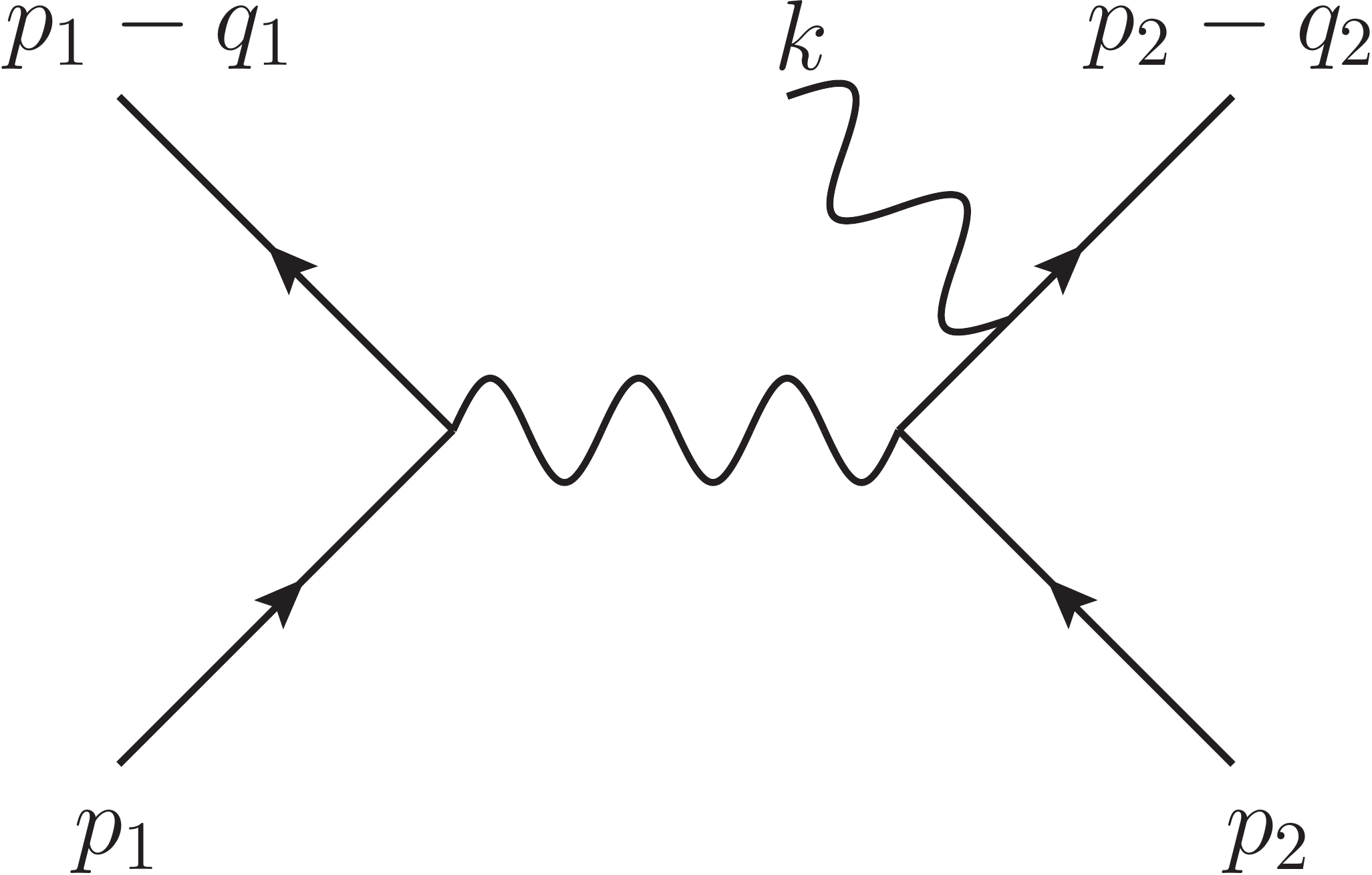}
		\caption{Diagram D}
		\label{fig:scalarDiagramsD}
	\end{subfigure}
	\begin{subfigure}[t]{0.32\textwidth}
		\includegraphics[width=0.9\textwidth]{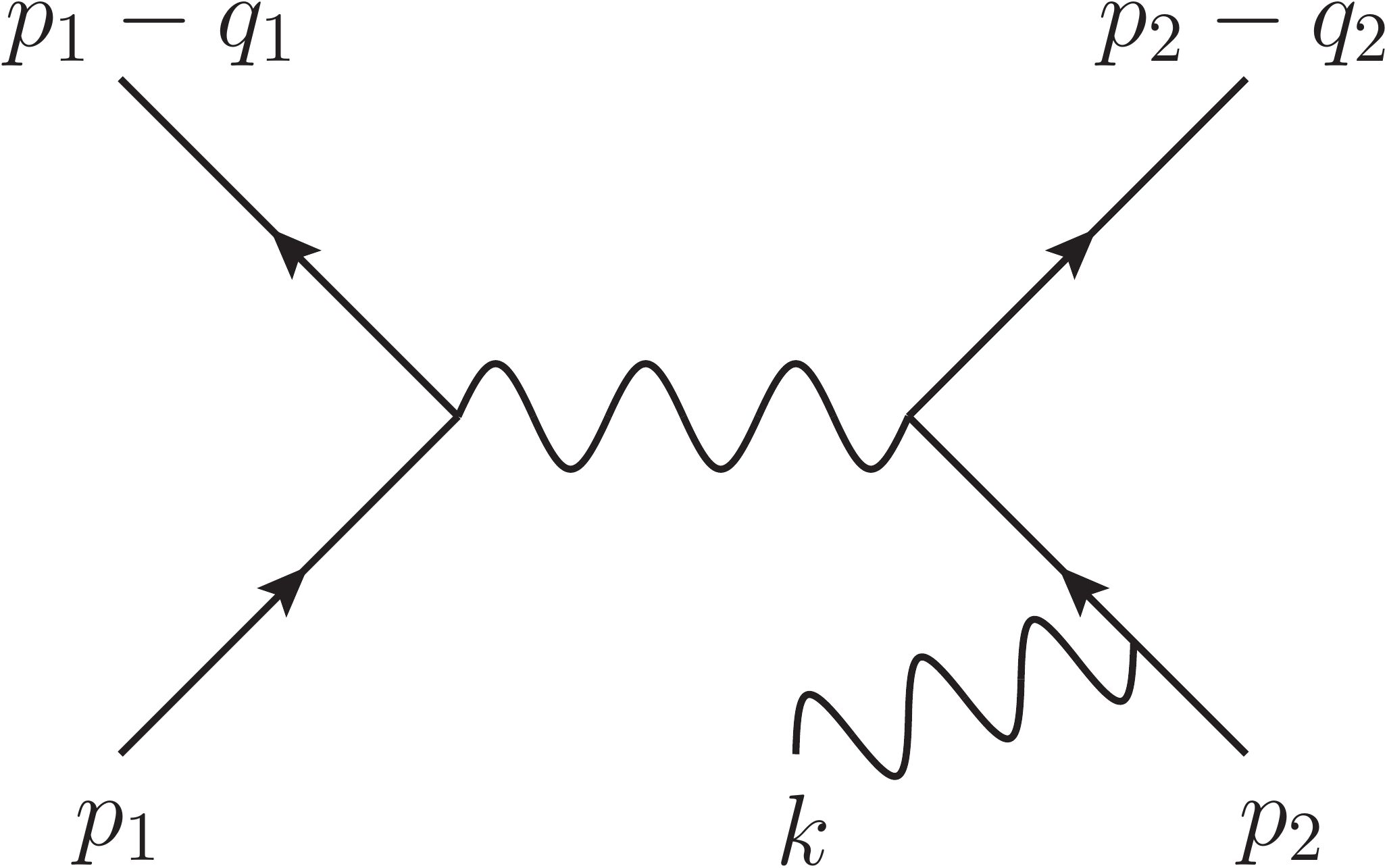}
		\caption{Diagram E}
		\label{fig:scalarDiagramsE}
	\end{subfigure}
	\caption{The five cubic diagrams for inelastic scalar
          scattering with gluon production in gauge theory, where time
          runs vertically.}
	\label{fig:scalarDiagrams}
\end{figure}

We have argued that the calculation of interest should be a tree
five-point amplitude. This is a very straightforward calculation using
Feynman diagrams. There are a total of seven Feynman diagrams, two of
which involve four point vertices. Thus there are five cubic diagrams,
shown in figure~\ref{fig:scalarDiagrams}, whose corresponding colour
factors are 
\begin{subequations}
\begin{align}
c_A &= (T_1^a \cdot T_1^b) T_2^b, \\
c_B &= (T_1^b \cdot T_1^a) T_2^b,\\
c_C &= f^{abc} T_1^b T_2^c, \\
c_D &= T_1^b (T_2^a \cdot T_2^b), \\
c_E &= T_1^b (T_2^b \cdot T_2^a).
\end{align}
\end{subequations}
The notation $T_1^a \cdot T_1^b$ indicates a matrix contraction of the group generators in representation $R_1$; $c_A$ is an element of the tensor product space $R_1 \otimes R_2$.

The total amplitude may then be written as
\begin{equation}
\mathcal{A} = \frac{n_A c_A}{d_A} + \frac{n_B c_B}{d_B} + \frac{n_C c_C}{d_C} + \frac{n_D c_D}{d_D} + \frac{n_E c_E}{d_E},
\end{equation}
where the kinematic numerators $n_i$ and the propagators $d_i$ can be
obtained by computing the seven Feynman diagrams, and assigning terms
from the four point vertices to cubic diagrams according to their
colour factors. Working in Feynman gauge\footnote{As they will not be relevant for our discussion, we omit factors of $i$ and couplings in our Feynman rules.}, the explicit results for the
numerators are:
\begin{subequations}
\label{eq:numerators}
\begin{align}
n_A &= (2 p_1 + q_2) \cdot (2 p_2 - q_2) \; \epsilon \cdot (2 p_1 + 2 q_2) - (2 p_1 \cdot q_2 + q_2^2) \; \epsilon \cdot(2 p_2 - q_2) , \\
n_B &= (2 p_1 - k - q_1) \cdot (2 p_2 - q_2) \; 2 \epsilon \cdot p_1 + 2 p_1 \cdot k \; \epsilon \cdot(2 p_2 - q_2),\\
n_C &= (2 p_1 - q_1)^\mu (2 p_2 - q_2)^\rho \left[(k+q_2)_\mu \eta_{\nu \rho} + (q_1 - q_2)_\nu \eta_{\rho\mu} - (k+q_1)_\rho \eta_{\mu\nu}\right] \epsilon^\nu, \\
n_D &= (2 p_1 - q_1) \cdot (2 p_2 + q_1) \; \epsilon \cdot (2 p_2 + 2 q_1) - (2 p_2 \cdot q_1 + q_1^2) \; \epsilon \cdot (2 p_1 - q_1),\\
n_E &= (2 p_1 - q_1) \cdot (2 p_2 -k - q_2) \;2 \epsilon \cdot p_2 + 2 p_2 \cdot k\; \epsilon \cdot (2 p_1 - q_1).
\end{align}
\end{subequations}
Notice that the symmetry of the situation requires that $n_D$ is simply equal to $n_A$ with particles 1 and 2 interchanged; similarly $n_E$ can be obtained from $n_B$. 

The propagators are straightforward to compute, yielding
\begin{subequations}
\label{eq:props}
\begin{align}
d_A &= (2 p_1 \cdot q_2 + q_2^2) \, q_2^2, \\
d_B &= - 2 p_1 \cdot k \, q_2^2,\\
d_C &=  q_1^2 \, q_2^2, \\
d_D &= (2 p_2 \cdot q_1 + q_1^2) \, q_1^2,\\
d_E &= -2 p_2 \cdot k \, q_1^2.
\end{align}
\end{subequations}
Once again, symmetry relates $d_A$ to $d_D$ and $d_B$ to $d_E$. From now on, we will exploit this symmetry and omit the explicit expressions for diagrams $D$ and $E$.

Before we construct the double copy, we must ensure that our numerators satisfy colour-kinematics duality; if they do not, we will need to modify them appropriately. In the case at hand, the colour factors satisfy precisely two identities:
\begin{equation}
c_A - c_B = c_C; \quad c_D - c_E = c_C.
\end{equation}
Therefore, we can construct a new gauge invariant amplitude via the double copy
\begin{equation}
\label{eq:gravAmp}
\mathcal{M} = \frac{n_A n_A}{d_A} + \frac{n_B n_B}{d_B} + \frac{n_C n_C}{d_C} + \frac{n_D n_D}{d_D} + \frac{n_E n_E}{d_E},
\end{equation}
if the kinematic numerators satisfy
\begin{equation}
\label{eq:kinematic}
n_A - n_B = n_C; \quad n_D - n_E = n_C.
\end{equation}
In fact, no
modification is necessary for this five-point tree level case: the
kinematic identities~\eqref{eq:kinematic} hold immediately in Feynman
gauge, as a direct calculation using the list of numerators in
equation~\eqref{eq:numerators} shows~\cite{Ochirov:2014gsa}. This favourable situation is not
expected to hold in general (i.e. for higher point amplitudes).

In view of the success of the double copy, we now have an expression, equation~\eqref{eq:gravAmp}, for a scattering amplitude in a gravitational theory. The motivation to construct this amplitude was to compare it to classical gravitational scattering: let us now see whether we have been successful.

\subsection{Large Mass Expansion \label{sec:ampToGR}}

In the previous section, we calculated an amplitude corresponding to
the scattering of two scalar particles, accompanied by the emission of
gravitational radiation. We would now like to compare this to the
classical scattering results in the Goldberger-Ridgway approach of
refs.~\cite{Goldberger:2016iau,Goldberger:2017frp}, reviewed here in
section~\ref{sec:goldberger}, and it is not immediately clear how the
two calculations are related. There are two issues to be
considered. The first is what constitutes classical
scattering. Generally accepted wisdom dictates that tree-level
diagrams correspond to classical physics, and loop diagrams provide
quantum corrections. However, there are subtleties in this argument,
as discussed in detail in ref.~\cite{Holstein:2004dn}, whose
conclusion is that loop integrals do indeed have a classical
component. To see why, one may consider the Lagrangian of our theory
with explicit factors of $\hbar$ reinstated:
\begin{equation}
\mathcal{L} = -\frac12 \tr F^{\mu\nu} F_{\mu\nu} + \left[\sum_i (D_\mu \Phi_i)^\dagger (D^\mu \Phi_i) - \frac{m^2_i}{\hbar^2} |\Phi|^2\right],
\end{equation}
where $D_\mu = \partial_\mu + i g A_\mu / \hbar$. In any given
amplitude, factors of $\hbar$ will occur associated with the couplings
and with the masses. Our amplitude is homogeneous in the couplings,
but not in the masses, so to take the classical limit we should treat
the mass $m_i$ as large. This is the source of classical corrections
from both tree and loop diagrams~\cite{Holstein:2004dn}\footnote{Note
  that the fact that classical corrections can come from either tree
  or loop diagrams also follows from the need for infrared
  singularities to cancel between real and virtual
  graphs~\cite{Bloch:1937pw,Kinoshita:1962ur,Lee:1964is}.}. Here, we
are requiring resolvable radiation in the final state, and studying
the lowest order Feynman diagrams for this to occur. Thus, we do not
need to add loop corrections to reproduce the results of
section~\ref{sec:goldberger}.

The second issue in relating the amplitude and equation of motion
approaches is the fact that the latter approach includes an expansion
in the deflections of the scattering particles. There are thus two
separate expansion parameters in principle: the coupling constant $g$, and the
momentum transfer between the scattering particles and the radiated
graviton, which measures the degree of deflection. As argued in reference~\cite{Goldberger:2016iau},
these expansions are correlated, in that the degree of deflection increases with each order of the coupling.
In our amplitude calculation, we therefore need to identify
the relevant expansion parameter that isolates this behaviour.

In fact, this idea already exists in
the literature. In particular, refs.~\cite{Laenen:2008gt,White:2011yy}
concerned the classification of radiation in both gauge theory and
gravity, up to and including ``next-to-soft'' terms in a systematic
expansion in the radiated momentum. The authors developed a physical
picture (based on the worldline formalism in quantum field
theory~\cite{Bern:1990cu,Bern:1991aq,Strassler:1992zr}), in which the
propagators for the scattering particles are replaced by quantum
mechanical (first-quantised) path integrals. These path integrals can
be calculated perturbatively, which corresponds to a sum over the
scattering particle trajectories, and thus the possible deflections of
the particles. Furthermore, the expansion of each path integral was
achieved by rescaling particle momenta according to $p^\mu\rightarrow
\lambda p^\mu$, before expanding in inverse power of $\lambda$. This
is precisely the large mass expansion alluded to above.

To extract the large masses, we will express the momenta of the incoming particles in terms of proper velocities, so $p_i^\mu = m_i v_i^\mu$. Then it is clear that the on-shell requirement for the incoming states translates to the statement that $v_i^2 = 1$. But we must also require that the outgoing states are on-shell, so
\begin{align}
&(p_i - q_i)^2 = m_i^2 - 2 m_i v_i \cdot q_i + q_i^2 = m_i^2\\
&\Rightarrow  2 m_i v_i \cdot q_i = q_i^2.
\end{align}
This equality is necessary to keep our amplitude on shell, so it is important to respect it scrupulously while performing the large mass expansion. Thus we treat the quantity $v_i \cdot q_i$ as of order $1/m$.

It will be useful to introduce some notation to keep track of various terms in the large mass expansion. The dominant terms in the list, equation~\eqref{eq:numerators}, of our numerators is of order $m^3$, and there are subleading corrections of order $m^2$ and lower. We will indicate this by writing $n_i = n_i^{(3)} + n_i^{(2)} + \cdots$. The dominant terms in the numerators are
\begin{subequations}
\label{eq:domNumerators}
\begin{align}
n_A^{(3)} &= 8 m_1^2 m_2 \;  v_1 \cdot v_2 \; \epsilon \cdot v_1 , \\
n_B^{(3)} &= 8 m_1^2 m_2 \;  v_1 \cdot v_2 \; \epsilon \cdot v_1  ,\\
n_C^{(3)} &= 0. 
\end{align}
\end{subequations}
We similarly expand the propagators $d_i$ in powers of the masses. They become
\begin{subequations}
\label{eq:domPropagators}
\begin{align}
d_A &= (2 p_1 \cdot q_2 + q_2^2) \, q_2^2 = 2 m_1 \, v_1 \cdot k \, q_2 ^2 + O(1/m), \\
d_B &= - 2 p_1 \cdot k \, q_2^2 = -2 m_1 \, v_1 \cdot k \, q_2^2, \\
d_C &=  q_1^2 \, q_2^2. 
\end{align}
\end{subequations}
We write these as $d_i = d_i^{(1)} + d_i^{(0)} + \cdots$.

At leading order in large masses, there is a considerable simplification. Since $n_A^{(3)} = n_B^{(3)}$ and $n_D^{(3)} = n_E^{(3)}$ to this order, while $d_A^{(1)} = -d_B^{(1)}$ and $d_D^{(1)} = - d_E^{(1)}$, it is easy to see that the dominant term in the gravitational amplitude vanishes. We need to go one order deeper in the large mass expansion to find anything interesting.

Straightforward Taylor expansions of the exact numerators in equation~\eqref{eq:numerators} lead to the next order corrections
\begin{subequations}
\begin{align}
n_A^{(2)} &= 8 m_1 m_2 \; v_1 \cdot v_2 \; \epsilon \cdot q_2 - 4 m_1 m_2 \; v_1 \cdot q_2 \; \epsilon \cdot v_2 - 4 m_1^2 \; \epsilon \cdot v_1 \; v_1 \cdot q_2 , \\
n_B^{(2)} &= 4 m_1 m_2 \left( v_1 \cdot k \; \epsilon \cdot v_2 - \epsilon \cdot v_1 \; v_2 \cdot k - \epsilon \cdot v_1 \; v_2 \cdot q_1 \right) - 4m_1^2 \; \epsilon \cdot v_1\;  v_1 \cdot q_2, \\
n_C^{(2)} &= 8 m_1 m_2 \left(v_1 \cdot q_2 \; \epsilon \cdot v_2 + q_1 \cdot \epsilon \; v_1 \cdot v_2 - v_2 \cdot q_1 \; v_1 \cdot \epsilon \right).
\end{align}
\end{subequations}
Meanwhile the corrections to the full propagators in equation~\eqref{eq:props} are
\begin{subequations}
\begin{align}
d_A^{(0)} &= q_2^2 - q_1^2, \\
d_B^{(0)} &= 0, \\
d_C^{(0)} &= d_C = q_1^2 q_2^2.
\end{align}
\end{subequations}
In terms of these quantities, the full gravitational amplitude is
\begin{equation}
\mathcal{M}_\textrm{cl} = - \frac{(n_A^{(3)})^2}{(d_A^{(1)})^2} d_A^{(0)} + 2\frac{n_A^{(3)} (n_A^{(2)} -  n_B^{(2)})}{d_A^{(1)}}+  \frac{(n_C^{(2)})^2}{d_C^{(0)}}  - \frac{(n_D^{(3)})^2}{(d_D^{(1)})^2} d_D^{(0)} + 2\frac{n_D^{(3)} (n_D^{(2)} -  n_E^{(2)})}{d_D^{(1)}}. 
\end{equation}
Notice that each term has a net four powers of mass. We have also written $\mathcal{M}_\textrm{cl}$ to indicate that this quantity is a classical limit of the tree amplitude.
After some algebra, the amplitude can be expressed in terms of the gauge-invariant $P$ and $Q$ vectors defined in equation~\eqref{eq:PQ}. The result is
\begin{multline}
\mathcal{M}_\textrm{cl} = 16 m_1^2 m_2^2 \; \epsilon_\mu \epsilon_\nu \left[
4 \frac{P_{12}^\mu P_{12}^\nu}{q_1^2 q_2^2} +2 \frac{v_1 \cdot v_2}{q_1^2 q_2^2} \left( Q_{12}^\mu P_{12}^\nu + Q_{12}^\nu P_{12}^\mu\right) \right.  \\
\left. + (v_1 \cdot v_2)^2 \left(\frac{Q_{12}^\mu Q_{12}^\nu}{q_1^2 q_2^2} - \frac{P_{12}^\mu P_{12}^\nu}{(k \cdot v_1)^2 (k\cdot v_2)^2} \right) 
\right].
\label{eq:gravAmplitude}
\end{multline}
It is instructive to compare this scattering amplitude against the expression, equation~\eqref{eq:classical}, for the classical radiation emitted during scattering. Evidently these quantities are closely related: the classical result of equation~\eqref{eq:classical} is an integral over the scattering amplitude in the large mass region, times certain factors.

To fully reconcile the classical calculation with the scattering amplitude in the large mass expansion, we need to bear in mind that, classically, the particles are associated with given states in position-space. We can write these as superpositions of momentum-space states as follows:
\begin{equation}
|\psi_i \rangle = \int \dd \! q_i \; \del(v_i \cdot q_i) e^{i q_i \cdot (b_i - x)} | q_i \rangle ,
\end{equation}
where $| q_i \rangle$ is a momentum eigenstate of the scalar field $\Phi_i$.
Working in the rest-frame of this particle, where $v_i = (1, 0, 0, 0)$, we can write the state as
\begin{equation}
|\psi_i \rangle = \int \frac{d^3q_i}{(2\pi)^3}  e^{-i \underline{q_i} \cdot (\underline{b_i} -\underline{x})} | q_i \rangle,
\end{equation}
which can be recognised as simply the wavefunction of a particle localised at the three-dimensional position $\underline{x} = \underline{b_i}$.

\section{Removing the Dilaton \label{sec:dilaton}}

The theory obtained via the double copy of pure Yang-Mills theory is
gravity, coupled to a dilaton and an antisymmetric tensor, the
axion. So far, we have been investigating a slightly different case:
the double copy of Yang-Mills theory coupled to a massive scalar. In
the double copy, we will therefore obtain gravity; a massless scalar
dilaton; a massless axion; and a massive scalar particle. The
graviton, dilaton and axion combine to make a single {\it product
  graviton} $H_{\mu\nu}$, and we must remove the scalar degrees of
freedom in order to obtain pure General Relativity. Given that all of
our solutions for $H_{\mu\nu}$ will be manifestly symmetric in the
indices $\mu$ and $\nu$, the axion never appears in what
follows. However, we must still construct a procedure for
removing the dilaton. We begin by investigating how the dilaton
couples to the massive scalar particles in the scattering process.

\subsection{Double Copy and Massive Amplitudes}

\begin{figure}[h]
\centering
	\includegraphics[height=0.2\textheight]{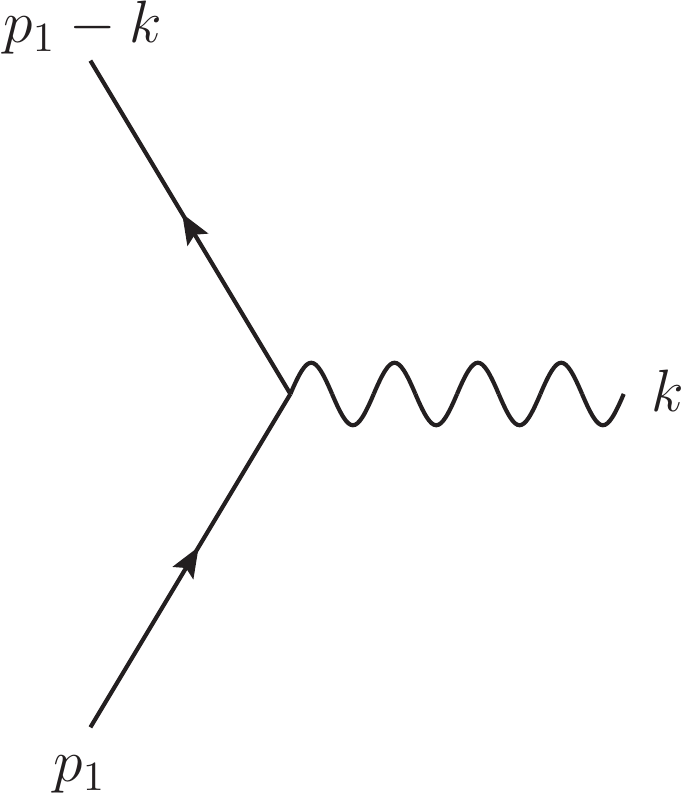}
	\caption{The three point interaction involving the massive gauge particle.}
	\label{fig:ym3pt}
\end{figure}

It is straightforward to construct three-point scattering amplitudes involving the massive particle. In the gauge theory, there is only the three point amplitude shown in figure~\ref{fig:ym3pt}, corresponding to a gauge field interacting with the scalar current.

The amplitude corresponding to this diagram is simply
\begin{equation}
\mathcal{A} = 2 \; p_1 \cdot \epsilon \; T_1,
\end{equation}
where $\epsilon$ is the polarisation vector associated with the vector particle. Notice that the requirement that all three external states are on-shell requires that $p_1 \cdot k = 0$.

Now we pass to the double-copied theory; we choose the polarisation vector in the second copy to be $\tilde \epsilon$. The amplitude is
\begin{equation}
\mathcal{M} = 4 \; p_1 \cdot \epsilon \; p_1 \cdot \tilde \epsilon.
\end{equation}
To determine how the massive particle couples to the dilaton, axion and graviton, we decompose the outer product of polarisation vectors into irreducible representations of the little group associated with the massless momentum $k$,
\begin{equation}
\epsilon^\mu \tilde{\epsilon}^\nu = \underbrace{\frac 12 \left(\epsilon^\mu \tilde{\epsilon}^\nu + \epsilon^\nu \tilde{\epsilon}^\mu - \frac2{d-2} \eta_\textrm{lg}^{\mu \nu} \right)}_\text{graviton} + \underbrace{\frac 12 \left(\epsilon^\mu \tilde{\epsilon}^\nu - \epsilon^\nu \tilde{\epsilon}^\mu \vphantom{\frac{2}{d-2}}\right)}_\text{axion} + \underbrace{\left(\frac1{d-2} \eta_\textrm{lg}^{\mu \nu}\right)}_\text{dilaton},
\label{eq:outerProduct}
\end{equation}
where 
\begin{equation}
\eta_\textrm{lg}^{\mu \nu} = \eta_{\mu\nu} - \frac{k_\mu q_\nu + k_\nu q_\mu}{k \cdot q},
\end{equation}
and $q$ is a choice of gauge satisfying $q \cdot q = 0$. 
The symmetry of $\mathcal{M}$ under interchanging $\epsilon$ and $\tilde \epsilon$ projects the antisymmetric tensor, corresponding to the axion, out. Meanwhile, the graviton and dilaton components of $\mathcal{M}$ are
\begin{align}
\mathcal{M}_\text{graviton} &= 4 p_1^\mu p_1^\nu e_{\mu\nu}, \\
\mathcal{M}_\text{dilaton} &= \frac{4}{d-2} m_1^2, \label{eq:dilatonCoupling}
\end{align}
where $e_{\mu\nu}$ is the traceless, symmetric polarisation tensor of the graviton.

A key point is that in the massive case $m_1 \neq 0$, there is a coupling between the massless dilaton and the massive scalar field. Correspondingly, a dilaton propagates in intermediate states in the five point amplitude we discussed previously in section~\ref{sec:gaugeScalarAmp}. Classically, the massive particles interact through a massless scalar force in addition to the gravitational force. While this is a natural feature of the double copy, it is desirable to be able to turn off this coupling to the dilaton: indeed, any application of the double copy to physical black hole scattering requires some means of disentangling contributions from dilatons. So let us now face this issue: how can we simply remove the dilaton diagrams from the double copy process?

\subsection{Dilatons in diagrams}

\begin{figure}[t]
\centering
	\begin{subfigure}[t]{0.3 \textwidth}
		\includegraphics[width=0.9\textwidth]{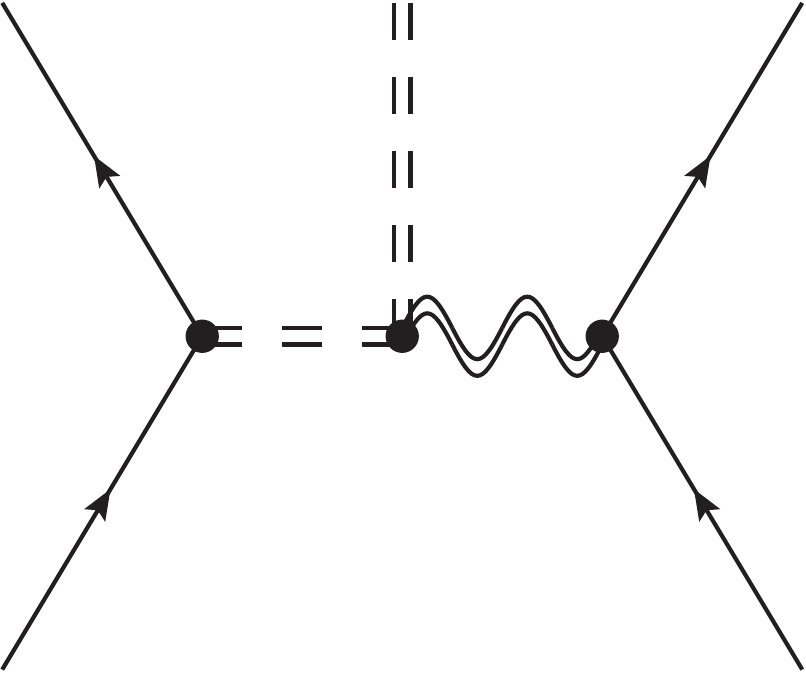}
	\end{subfigure}
	\begin{subfigure}[t]{0.3 \textwidth}
		\includegraphics[width=0.9\textwidth]{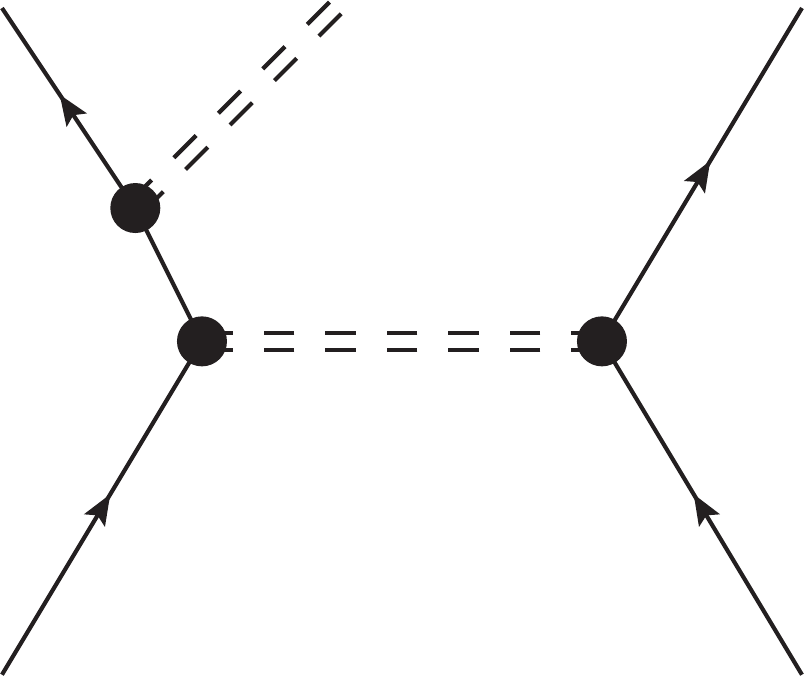}
	\end{subfigure}
	\caption{Diagrams with dilatons radiated into the final state. The dilaton is represented by the double dashed line, while the graviton is the double wavy line.}
	\label{fig:finalDilaton}
\end{figure}

\begin{figure}[t]
\centering
	\begin{subfigure}[t]{0.3 \textwidth}
		\includegraphics[width=0.9\textwidth]{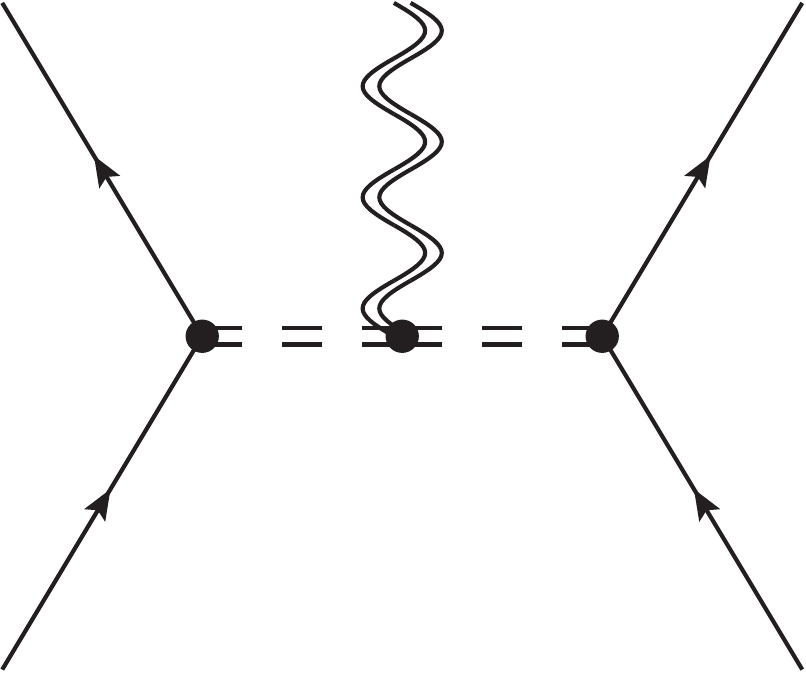}
	\end{subfigure}
	\begin{subfigure}[t]{0.3 \textwidth}
		\includegraphics[width=0.9\textwidth]{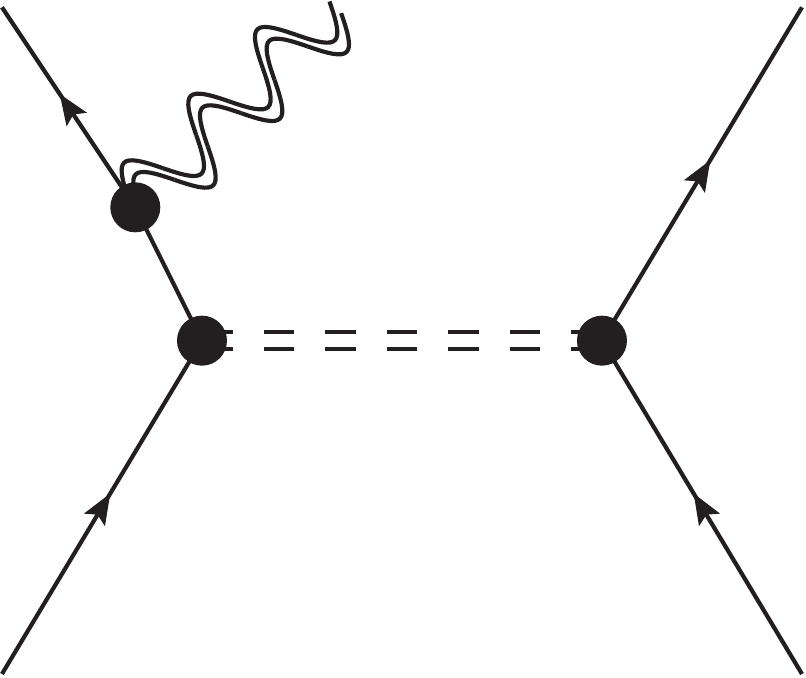}
	\end{subfigure}
	\caption{Types of diagram with only intermediate, virtual, dilaton states.}
	\label{fig:toRemove}
\end{figure}

It is generally straightforward to identify the contributions of particular substates in scattering amplitudes by looking at their cuts or factorisation channels. In the present case, we already know how the dilaton appears in the five point amplitude. It is convenient to consider two categories of diagram involving the dilaton: those with external dilatons, for example the diagrams in figure~\ref{fig:finalDilaton}, and those without any external dilatons. Examples of diagrams in this second class are shown in figure~\ref{fig:toRemove}. Of course dilatons may be present as virtual states in both categories. 

The first class of diagram, with dilatons in the final state, is trivial to deal with. We simply project our amplitude onto the traceless symmetric component by hand by replacing $\epsilon^\mu \tilde \epsilon^\nu \rightarrow e^{\mu\nu}$. In doing so, we have thrown away the dilaton polarisation tensor which is the trace term present in the tensor decomposition of the outer product of polarisation vectors, equation~\eqref{eq:outerProduct}. This replacement explicitly forces the external state to be a graviton, rather than a dilaton -- thus removing all diagrams in our first category.
But this procedure does not remove the diagrams in our second category, in which virtual dilatons may be present.

A few methods for removing virtual dilatons suggest themselves. The first method, advocated in~\cite{Luna:2016hge}, is to insert a projector onto graviton states at vertices in which the massive scalar line may interact with a dilaton. A second, related, option is to insert a projector on the intermediate lines. But there is also a third option which is very simple to implement. We observe that the dilaton is a scalar particle propagating in the double copy. It is possible to remove such scalar particles by introducing a ghost: another scalar particle in the double copy, but where the double copy is defined by introducing a negative sign any time this particle appears. This method has been used in the context of the double copy by Johansson and Ochirov~\cite{Johansson:2014zca} to obtain pure gravity as a double copy.

We will therefore introduce a new massless scalar state in our gauge theory. We can constrain how the new scalar behaves by requiring that it removes the coupling between our massive scalar and the dilaton in the double copy. In particular, the ghost must remove the diagrams in figure~\ref{fig:toRemove}. We will therefore assume that the ghost couples to two massive scalars, and that the ghost is charged under the gauge symmetry, transforming in the adjoint representation. The Lagrangian of the theory is
\begin{equation}
\mathcal{L} = -\frac 12 \tr F_{\mu\nu} F^{\mu\nu} +\tr D^\mu \chi D_\mu \chi + \sum_i \left[(D_\mu \Phi_i)^\dagger D^\mu \Phi_i - m_i^2 \Phi_i^\dagger \Phi_i  - 2 X m_i \Phi_i^\dagger \chi \Phi_i \right],
\label{eq:extendedGauge}
\end{equation}
where $\chi$ is the adjoint ghost, and $X$ is a coupling to be determined\footnote{The factor of $m_i$ in the coupling is inserted so that $X$ is dimensionless, and using the knowledge from equation~\eqref{eq:dilatonCoupling} that the dilaton couples to mass.}.

\subsection{Example at Four Points}

\begin{figure}[h]
\centering
	\begin{subfigure}[t]{0.3\textwidth}
		\includegraphics[width=0.9\textwidth]{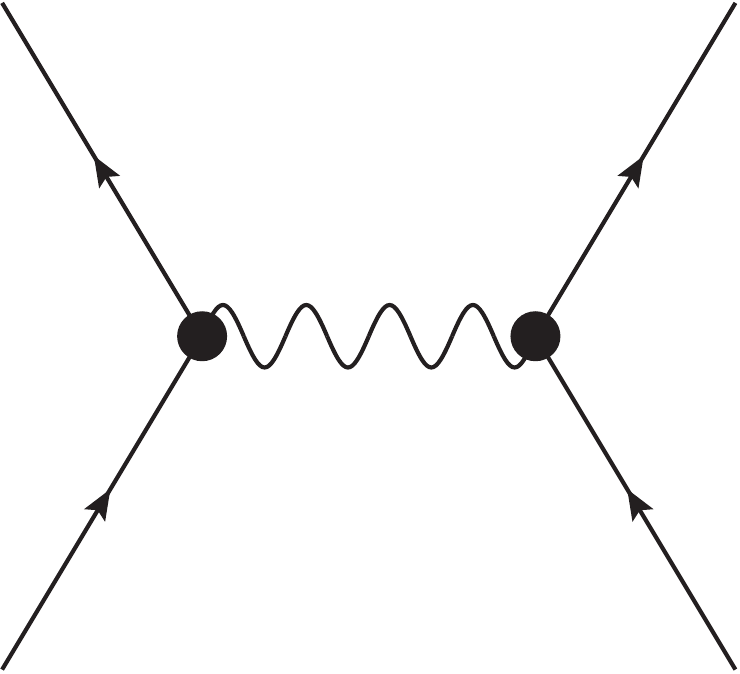}
		\caption{Gauge boson mediator}
	\end{subfigure}
	\begin{subfigure}[t]{0.3\textwidth}
		\includegraphics[width=0.9\textwidth]{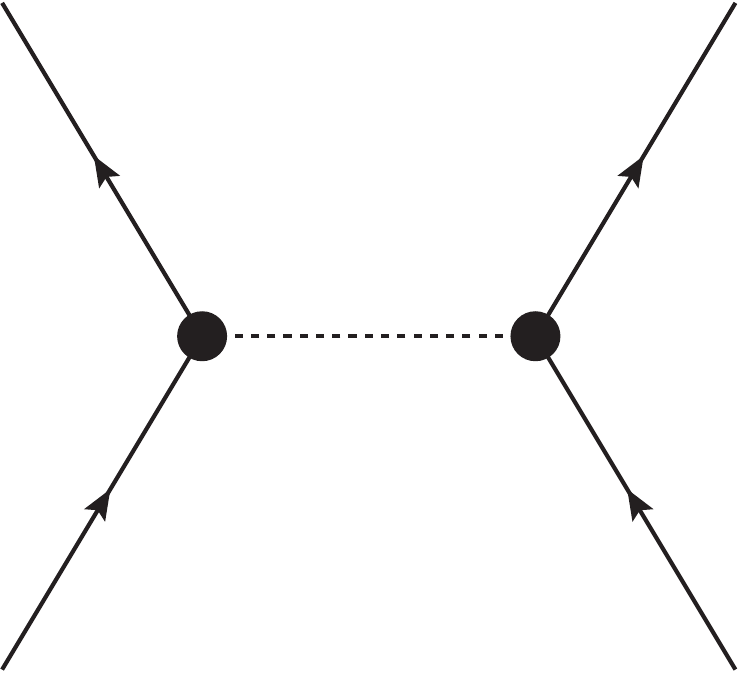}
		\caption{Ghost mediator}
	\end{subfigure}
	\caption{Four point scattering in the extended gauge theory of equation~\eqref{eq:extendedGauge}.}
	\label{fig:fourPtDiags}
\end{figure}

To see how the procedure works in the simplest case, we turn to elastic two-particle scattering. We will determine the amplitude for two massive scalar particles scattering off one another in General Relativity from the double copy. Beginning in gauge theory, there are only two diagrams to compute, as shown in figure~\ref{fig:fourPtDiags}. 

We denote the incoming momenta by $p_1$ and $p_2$ and let the momentum transfer be $q$. As we will treat the gauge and ghost mediator cases slightly differently, we present the contributions separately; they are
\begin{align}
\mathcal{A}_\text{gauge} = \frac{4 p_1 \cdot p_2 + q^2}{q^2} T_1 \cdot T_2, \\
\mathcal{A}_\text{ghost} =  X^2 \frac{4 m_1 m_2}{q^2} T_1 \cdot T_2. 
\end{align}
At this order, colour-kinematics duality is trivial so the double copy is immediate. We define the double copy for the ghost by inserting a sign, so that the gravitational amplitude is
\begin{equation}
\mathcal{M} = \frac{1}{q^2} \left[ (4 p_1 \cdot p_2 + q^2)^2 - X^4 (4 m_1 m_2)^2 \right].
\end{equation}
Now, on the factorisation channel where $q^2 = 0$, the quantity $q^2 \mathcal{M}$ must factorize into a product of three point amplitudes, summed over intermediate helicities. We find
\begin{align}
q^2 \mathcal{M} &\xrightarrow[q^2 \rightarrow 0]{} (4 p_1 \cdot p_2)^2 - X^4 (4 m_1 m_2)^2 \\
&= 8 p_1^\mu p_1^{\mu'} p_2^\nu p_2^{\nu'} \left[ \eta_{\mu \nu} \eta_{\mu' \nu'} +\eta_{\mu \nu'} \eta_{\mu' \nu} - 2X^4 \eta_{\mu \mu'} \eta_{\nu \nu'} \right].
\end{align}
In General Relativity, the quantity in square brackets in the last line must be equal to the de Donder projector
\begin{equation}
P_{\mu \mu' \; \nu \nu'} = \left[ \eta_{\mu \nu} \eta_{\mu' \nu'} +\eta_{\mu \nu'} \eta_{\mu' \nu} - \frac{2}{d-2} \eta_{\mu \mu'} \eta_{\nu \nu'} \right],
\end{equation}
up to pure gauge terms. We therefore conclude that 
\begin{equation}
X^4 = \frac{1}{d-2}.
\label{eq:valueOfX}
\end{equation}

\subsection{Inelastic scattering}
Our next task is to re-compute the five-point scattering amplitude in gauge theory, including the new state. It is convenient to use the same cubic topologies as presented in figure~\ref{fig:scalarDiagrams}. The contributions to the amplitude due to the presence of the $\chi$ are given by
\begin{eqnarray}
\label{eq:ghostAmp}
\mathcal{A}_{\textrm{\tiny{ghost}}} = \frac{c_A n'_A}{d_A} + \frac{c_B n'_B}{d_B} + \frac{c_C n'_C}{d_C} + \frac{c_D n'_D}{d_D} + \frac{c_E n'_E}{d_E}.
\end{eqnarray}
A straightforward calculation yields the new terms in the numerators $n'_A, n'_B$ and $n'_C$. These new terms are
\begin{subequations}
\label{eq:chiNumerators}
\begin{align}
n'_A &= 4 m_1 m_2 X^2 \; 2 \epsilon \cdot (p_1 + q_2), \\
n'_B &= 4 m_1 m_2 X^2 \; 2 \epsilon \cdot p_1 ,\\
n'_C &= -4 m_1 m_2 X^2 \; \epsilon \cdot (q_1 - q_2) .
\end{align}
\end{subequations}
The numerators to $n'_D$ and $n'_E$ can be obtained by swapping the particle labels 1 and 2 in $n'_A$ and $n'_B$ as before. 
It is easy to see that the these numerators satisfy the relation
\begin{equation}
n'_A - n'_B = n'_C,
\end{equation}
so that colour-kinematics duality is satisfied, and we can construct the double copy as before. The contribution of the new state to the amplitude is
\begin{eqnarray}
\label{eq:gravAmpGRscalar}
\mathcal{M}_{\textrm{\tiny{scalar}}} = \frac{n'_A n'_A}{d_A} + \frac{n'_B n'_B}{d_B} + \frac{n'_C n'_C}{d_C} + \frac{n'_D n'_D}{d_D} + \frac{n'_E n'_E}{d_E}.
\end{eqnarray}
Inserting the explicit expressions, and replacing the product of polarisation vectors $\epsilon_\mu \epsilon_\nu$ by the traceless symmetric graviton polarisation tensor $e_{\mu\nu}$ to remove final state dilatons, we find the explicit amplitude
\begin{eqnarray}
\mathcal{M}_{\textrm{\tiny{scalar},cl}} = 16 m_1^2 m_2^2 \; e_{\mu \nu} \left[
\frac{1}{d-2} \left(\frac{Q_{12}^\mu Q_{12}^\nu}{q_1^2 q_2^2} - \frac{P_{12}^\mu P_{12}^\nu}{(k \cdot v_1)^2 (k\cdot v_2)^2} \right) 
\right],
\label{eq:gravAmpGRscalar2}
\end{eqnarray}
where we used the value of $X$, equation~\eqref{eq:valueOfX}, determined at four points, and we have performed a large mass expansion. The Einstein gravity amplitude is then easily obtained by subtracting the scalar contributions from the gravitational amplitude computed in the previous section. This is
\begin{equation}
\label{eq:gravAmpGR}
\mathcal{M}_{\textrm{\tiny{GR}}} = 
\mathcal{M_\textrm{cl}}-
\mathcal{M}_{\textrm{\tiny{scalar},cl}},
\end{equation}
where $\mathcal{M}_\textrm{cl}$ is defined as in equation~\eqref{eq:gravAmplitude}. The Einstein gravity amplitude takes the explicit form
\begin{multline}
\mathcal{M}_{\textrm{\tiny{GR}}} = 16 m_1^2 m_2^2 \; e_{\mu \nu} \left[
4 \frac{P_{12}^\mu P_{12}^\nu}{q_1^2 q_2^2} +2 \frac{v_1 \cdot v_2}{q_1^2 q_2^2} \left( Q_{12}^\mu P_{12}^\nu + Q_{12}^\nu P_{12}^\mu\right) \right.  \\
\left. + \left((v_1 \cdot v_2)^2 - \frac{1}{d-2} \right) \left(\frac{Q_{12}^\mu Q_{12}^\nu}{q_1^2 q_2^2} - \frac{P_{12}^\mu P_{12}^\nu}{(k \cdot v_1)^2 (k\cdot v_2)^2} \right) 
\right].
\end{multline}
The classical graviton associated with this amplitude can therefore be taken to be
\begin{multline}
k^2 h^{(1)\mu\nu}(k) = - \frac{m_1 m_2}{8 m_\textrm{pl}^{3(d-2)/2}} \int \dd q_1 \dd q_2 \del(k - q_1 - q_2) \del(q_1 \cdot v_1) e^{i q_1 \cdot b_1}  \del(q_2 \cdot v_2) e^{i q_2 \cdot b_2} \times \\
\left[
\frac{P_{12}^\mu P_{12}^\nu}{q_1^2 q_2^2} +\frac{v_1 \cdot v_2}{2q_1^2 q_2^2} \left( Q_{12}^\mu P_{12}^\nu + Q_{12}^\nu P_{12}^\mu\right) 
+ \frac14 \left((v_1 \cdot v_2)^2 - \frac{1}{d-2} \right)  \times
\right. \\ \left. 
\left(\frac{Q_{12}^\mu Q_{12}^\nu}{q_1^2 q_2^2} - \frac{P_{12}^\mu P_{12}^\nu}{(k \cdot v_1)^2 (k\cdot v_2)^2} \right) 
\right],
\end{multline}
up to pure gauge terms. We have checked that this result is accurate by comparison with a far more complicated direct computation in General Relativity coupled to point particles.

\subsection{Relation to classical field computation}

Given that our treatment was motivated by the computation performed by Goldberger and Ridgway, it is fair to ask if the method we employed to remove the dilaton can also be implemented in their framework. We will consider an analogue of the classical gauge theory equations \eqref{eq:classicalEqns} for a ghost field $\chi$ that couples to the classical coloured point particles as well as the gluons. The equations defining the contributions from the ghost to the deflection of the colour charged point particles are
\begin{subequations}
\begin{align}
\partial^2\chi^a &=  \sum_i 2Xgm_i\int d \tau c_i^a(\tau) \delta^{(d)}(x - x_i(\tau)), \\
m_i \frac{d^2 x_i^\mu(\tau)}{d \tau^2} &= 2Xgm_i \partial^\mu \chi^a(x_i(\tau)) c^a_i(\tau), 
\\
\frac{dc_i^a(\tau)}{d\tau} &= 2Xgm_i f^{abc}\chi^b(x_i(\tau)) \, c^c_i(\tau).
\end{align}
\end{subequations}
We can perform now an analogous computation to the one used to get equation~\eqref{eq:classicalA}. This will yield the contributions to the gluon field due to the presence of the ghost field $\chi$, and it takes the explicit form 
\begin{multline}
k^2 A^{(1)a\mu}(k)\Big|_{\textrm{ghost}} = g^3  \int \frac{\dd q_1 \dd q_2}{q_2^2} \del(k - q_1 - q_2) \del(q_1 \cdot v_1) e^{i q_1 \cdot b_1}  \del(q_2 \cdot v_2) e^{i q_2 \cdot b_2}  \\
\times 2m_12m_2X^2\frac{1}{k \cdot p_1} \left[if^{abc} c_2^{(0)b} c_1^{(0)c} p_1^{\mu}
+c_1^{(0)a} c_1^{(0)} \cdot c_2^{(0)} \left(q_2^\mu 
+ \frac{p_1^\mu}{k \cdot p_1} k \cdot q_2\right)\right.\\ -if^{abc} c_2^{(0)b} c_1^{(0)c} \frac{(q_1^{\mu}-q_2^{\mu})}{q_1^2}k \cdot p_1  \bigg]  + (1 \leftrightarrow 2).
\label{eq:classicalghost}
\end{multline}
If we consider the leading terms in a large mass expansion of the ghost contributions to the scattering amplitude in equation~\eqref{eq:ghostAmp}: 
\begin{align}
\label{eq:ghostlargemass1}\frac{c_A n'_A}{d_A}&\rightarrow 4 m_1 m_2 X^2 c_A
\left(\frac{\epsilon\cdot p_1}{p_1\cdot k}
+\frac{\epsilon\cdot q_2}{p_1\cdot k}
+\frac{\epsilon\cdot p_1\ q_2\cdot k}{(p_1\cdot k)^2}
\right)\\
\label{eq:ghostlargemass2}\frac{c_B n'_B}{d_B}&\rightarrow -4 m_1 m_2 X^2 c_B\frac{\epsilon\cdot p_1}{p_1\cdot k}\\
\label{eq:ghostlargemass3}\frac{c_C n'_C}{d_C}&\rightarrow -4 m_1 m_2 X^2 c_C\frac{\epsilon \cdot (q_1 - q_2)}{q_1^2q_2^2},
\end{align}
it is easy to see that the equations~\eqref{eq:ghostlargemass1} and~\eqref{eq:ghostlargemass2} combine to yield the first terms of equation~\eqref{eq:classicalghost}, while the contributions from equation~\eqref{eq:ghostlargemass3} are directly responsible for the final line of equation~\eqref{eq:classicalghost}. Thus we see that we can indeed remove the dilaton pollution, this time completely within the framework of classical perturbation theory.

This result shows a direct link between a classical computation
similar to that of refs. \cite{Goldberger:2016iau,Goldberger:2017frp}
and an amplitude, so we may ask if the double copy (scalar) scattering
amplitude from equation~\eqref{eq:gravAmpGRscalar2} can be obtained
using a set of replacements similar to those proposed
in~\cite{Goldberger:2016iau}:
\begin{align}
c_i^{(0)a} &\rightarrow m_i v_i^{\mu} ,\\
if^{abc} &\rightarrow \frac12 \Gamma^{\mu\nu\rho}(q_1, q_2, q_3 ).
\label{eq:replacementgluon2}
\end{align}
It is not difficult to see that such double copy rules will not land on the amplitude from equation~\eqref{eq:gravAmpGRscalar2}. Instead, it is the set of replacements 
\begin{align}
c_1^{(0)a} c_1^{(0)} \cdot c_2^{(0)} &\rightarrow 2m_12m_2X^2p_1^{\mu} ,\\
if^{abc} c_2^{(0)b} c_1^{(0)c} &\rightarrow 2m_12m_2X^2\frac12 (q_1^\mu-q_2^\mu),
\label{eq:replacementghost2}
\end{align} 
that one needs to obtain the desired amplitude. Of course, this is because the process we are considering now depends on the dynamics of the ghost field, and so it should be no surprise that the replacement in equation~\eqref{eq:replacementghost2} involves the ghost-gluon vertex instead of the three gluon vertex of equation~\eqref{eq:replacementgluon2}. Nevertheless, this highlights one advantage of the scattering amplitudes: when they are available, it is straightforward to take the double copy. But in situations where scattering amplitudes are not available, it could well be that developing these replacement rules, and supplementing them by some notion of colour-kinematics duality, could allow the double copy to access entirely new physical regimes.

\section{Discussion and Conclusions}\label{sec:conclusions}

We hope this work represents a step towards using the double copy as a tool for understanding the classical physics of General Relativity. Future gravitational wave observatories, such as LISA or the Einstein telescope, will operate at higher precision, so the demand for understanding the finest details of gravity will become more pressing. Since precision has always been a driving force in the study of scattering amplitudes, it is natural to investigate whether amplitudes methods have relevance for precision General Relativity.

We began by reformulating the leading perturbative term in the inelastic gravitational scattering of two singularities as a tree scattering amplitude, following Goldberger and Ridgway's double-copy based calculation. This reformulation allowed us to analyze the factorisation structure of the calculation, and to identify one method of removing the dilaton field which was present in the original calculation. We did so by introducing a new scalar field in the gauge theory which is double-copied to gravity. This scalar is treated as a ghost, in a manner similar to Johansson and Ochirov's construction~\cite{Johansson:2014zca} of loop amplitudes in pure gravity. However, we have not shown that our method will work at higher orders. Indeed it is likely that there will be issues, since ghosts built from double copies of scalars (rather than double copies of spinors) encounter problems at two loops~\cite{Johansson:2014zca}.
Our attitude is that an understanding of how best to remove the dilaton will depend on how one computes higher order corrections to the classical scattering process, which we leave for future work.

An interesting feature of Goldberger and Ridgway's work was an unusual
implementation of the double copy,
equation~\eqref{eg:grReplacement}. Essentially, they replaced the
colour structure constants by the kinematic Yang-Mills three point
vertex, and the colour factors by the momenta. This is a little
puzzling, because one would expect that some work should be done to
synchronise the colour and kinematic structure in the calculation. In
our amplitude-based approach, the standard BCJ replacement of colour
factors by kinematic numerators was available. But we also encountered
a simplification: colour-kinematics duality holds for free (in Feynman
gauge.) This simplification is most unlikely to pertain to higher
orders, so we anticipate that the replacements Goldberger and Ridgway
performed will at least need to be supplemented by some kind of
colour-kinematics duality condition in more general cases. On a
related note, it was satisfying to see a version of the
Goldberger-Ridgway replacement appear in the context of our scalar
double copy, equation~\eqref{eq:replacementghost2}. It could well be
that in situations where scattering amplitudes are unavailable, a
version of the double copy based on these kinds of replacement rules
could still work. This may allow the double copy to be used in
entirely new ways.

We believe that there is considerable scope for further work on applying the double copy in the context of classical General Relativity. There is obvious motivation to pursue this work to higher orders in perturbation theory, in the cases of elastic and inelastic scattering. It will also be important to determine how to handle the angular momentum of a black hole. In this work, we were insensitive to black hole spin, but higher order corrections will probe this aspect. One issue that is likely to arise is another kind of unwanted state propagating in the double copy: the axion. The axion cannot couple to a spinless point particle, but when a non-trivial spin vector is present a coupling exists. Therefore it is likely to be necessary to remove propagating axions; this is an issue which has already been addressed in loop calculations~\cite{Johansson:2014zca}. This could be an additional complication, but nevertheless it is worth noticing recent, encouraging, progress in capturing spin effects in classical GR~\cite{Guevara:2017csg,Vines:2017hyw}.

Finally, our goal has been to develop the double copy as a tool to simplify perturbation theory in General Relativity. But the double copy is not the only new idea for simplifying GR. It could be that a better approach~\cite{Bern:1999ji,Cheung:2016say,Cheung:2017kzx} is to design a new Lagrangian for GR, equivalent to the Einstein-Hilbert Lagrangian, but exhibiting a simpler perturbative expansion. This may lead to a simpler algorithm for determining precision gravitational effects. Such a procedure still has the double copy at its heart, however, and thus we expect the double copy to play a key role in gravitational perturbation theory in the coming years.

\begin{acknowledgments}
We thank Zvi Bern, Marco Chiodaroli, Walter Goldberger, Enrico Herrmann, David Kosower, Ricardo Monteiro, Alec Ridgway, Julio Parra-Martinez and Radu Roiban for helpful discussions.
AL is funded by a
Conacyt studentship, and thanks the Higgs Centre for
hospitality. 
DOC is an IPPP associate, and thanks the IPPP for on-going support
as well as for hospitality during this work. He is supported in part by the Marie Curie FP7 grant 631370 and by the STFC consolidated grant ``Particle Physics at the Higgs Centre''.
IN is supported by STFC studentship ST/N504051/1.
Meanwhile CDW is supported by the UK Science and Technology Facilities Council (STFC). This research was supported in part by the National Science Foundation under Grant No. NSF PHY-1125915.
\end{acknowledgments}

\bibliographystyle{JHEP}

\begin{thebibliography}{99}

\bibitem{Kawai:1985xq} 
  H.~Kawai, D.~C.~Lewellen and S.~H.~H.~Tye,
  Nucl.\ Phys.\ B {\bf 269}, 1 (1986).
  doi:10.1016/0550-3213(86)90362-7

\bibitem{Bern:2008qj} 
  Z.~Bern, J.~J.~M.~Carrasco and H.~Johansson,
  Phys.\ Rev.\ D {\bf 78}, 085011 (2008)
  doi:10.1103/PhysRevD.78.085011
  [arXiv:0805.3993 [hep-ph]].
  
\bibitem{Bern:2010ue} 
  Z.~Bern, J.~J.~M.~Carrasco and H.~Johansson,
  Phys.\ Rev.\ Lett.\  {\bf 105}, 061602 (2010)
  doi:10.1103/PhysRevLett.105.061602
  [arXiv:1004.0476 [hep-th]].

\bibitem{Neill:2013wsa} 
  D.~Neill and I.~Z.~Rothstein,
  Nucl.\ Phys.\ B {\bf 877}, 177 (2013)
  doi:10.1016/j.nuclphysb.2013.09.007
  [arXiv:1304.7263 [hep-th]].

\bibitem{Bjerrum-Bohr:2013bxa} 
  N.~E.~J.~Bjerrum-Bohr, J.~F.~Donoghue and P.~Vanhove,
  JHEP {\bf 1402}, 111 (2014)
  doi:10.1007/JHEP02(2014)111
  [arXiv:1309.0804 [hep-th]].

\bibitem{Bjerrum-Bohr:2014lea} 
  N.~E.~J.~Bjerrum-Bohr, B.~R.~Holstein, L.~Plant\'e and P.~Vanhove,
  Phys.\ Rev.\ D {\bf 91}, no. 6, 064008 (2015)
  doi:10.1103/PhysRevD.91.064008
  [arXiv:1410.4148 [gr-qc]].

\bibitem{Bjerrum-Bohr:2014zsa} 
  N.~E.~J.~Bjerrum-Bohr, J.~F.~Donoghue, B.~R.~Holstein, L.~Plant\'e and P.~Vanhove,
  Phys.\ Rev.\ Lett.\  {\bf 114}, no. 6, 061301 (2015)
  doi:10.1103/PhysRevLett.114.061301
  [arXiv:1410.7590 [hep-th]].
  
 
\bibitem{Bjerrum-Bohr:2016hpa} 
  N.~E.~J.~Bjerrum-Bohr, J.~F.~Donoghue, B.~R.~Holstein, L.~Plante and P.~Vanhove,
  JHEP {\bf 1611}, 117 (2016)
  doi:10.1007/JHEP11(2016)117
  [arXiv:1609.07477 [hep-th]].

\bibitem{Cachazo:2017jef} 
  F.~Cachazo and A.~Guevara,
  arXiv:1705.10262 [hep-th].
  
\bibitem{Guevara:2017csg} 
  A.~Guevara,
  arXiv:1706.02314 [hep-th].

\bibitem{Bjerrum-Bohr:2017dxw} 
  N.~E.~J.~Bjerrum-Bohr, B.~R.~Holstein, J.~F.~Donoghue, L.~Plant\'e and P.~Vanhove,
  PoS CORFU {\bf 2016}, 077 (2017)
  [arXiv:1704.01624 [gr-qc]].

\bibitem{Damour:2017zjx} 
  T.~Damour,
  arXiv:1710.10599 [gr-qc].

\bibitem{Monteiro:2014cda} 
  R.~Monteiro, D.~O'Connell and C.~D.~White,
  JHEP {\bf 1412}, 056 (2014)
  doi:10.1007/JHEP12(2014)056
  [arXiv:1410.0239 [hep-th]].

\bibitem{Luna:2015paa} 
  A.~Luna, R.~Monteiro, D.~O'Connell and C.~D.~White,
  Phys.\ Lett.\ B {\bf 750}, 272 (2015)
  doi:10.1016/j.physletb.2015.09.021
  [arXiv:1507.01869 [hep-th]].

\bibitem{Ridgway:2015fdl} 
  A.~K.~Ridgway and M.~B.~Wise,
  Phys.\ Rev.\ D {\bf 94}, no. 4, 044023 (2016)
  doi:10.1103/PhysRevD.94.044023
  [arXiv:1512.02243 [hep-th]].

\bibitem{Luna:2016due} 
  A.~Luna, R.~Monteiro, I.~Nicholson, D.~O'Connell and C.~D.~White,
  JHEP {\bf 1606}, 023 (2016)
  doi:10.1007/JHEP06(2016)023
  [arXiv:1603.05737 [hep-th]].
  
\bibitem{Adamo:2017nia}
  T.~Adamo, E.~Casali, L.~Mason and S.~Nekovar,
  arXiv:1706.08925 [hep-th].

\bibitem{Adamo:2017sze}
  T.~Adamo, E.~Casali, L.~Mason and S.~Nekovar,
  arXiv:1708.09249 [hep-th].

\bibitem{Bahjat-Abbas:2017htu} 
  N.~Bahjat-Abbas, A.~Luna and C.~D.~White,
  arXiv:1710.01953 [hep-th].
  
\bibitem{Carrillo-Gonzalez:2017iyj} 
  M.~Carrillo-Gonzalez, R.~Penco and M.~Trodden,
  arXiv:1711.01296 [hep-th].

\bibitem{Luna:2016hge}
  A.~Luna, R.~Monteiro, I.~Nicholson, A.~Ochirov, D.~O'Connell, N.~Westerberg and C.~D.~White,
  JHEP {\bf 1704} (2017) 069
  doi:10.1007/JHEP04(2017)069
  [arXiv:1611.07508 [hep-th]].

\bibitem{Goldberger:2016iau}
  W.~D.~Goldberger and A.~K.~Ridgway,
  Phys.\ Rev.\ D {\bf 95} (2017) no.12,  125010
  doi:10.1103/PhysRevD.95.125010
  [arXiv:1611.03493 [hep-th]].
  
\bibitem{Goldberger:2017frp}
  W.~D.~Goldberger, S.~G.~Prabhu and J.~O.~Thompson,
  Phys.\ Rev.\ D {\bf 96} (2017) no.6,  065009
  doi:10.1103/PhysRevD.96.065009
  [arXiv:1705.09263 [hep-th]].
  
\bibitem{Janis:1968zz} 
  A.~I.~Janis, E.~T.~Newman and J.~Winicour,
  Phys.\ Rev.\ Lett.\  {\bf 20}, 878 (1968).
  doi:10.1103/PhysRevLett.20.878
  
\bibitem{Bern:2010yg} 
  Z.~Bern, T.~Dennen, Y.~t.~Huang and M.~Kiermaier,
  Phys.\ Rev.\ D {\bf 82}, 065003 (2010)
  doi:10.1103/PhysRevD.82.065003
  [arXiv:1004.0693 [hep-th]].
  
\bibitem{Du:2011js} 
  Y.~J.~Du, B.~Feng and C.~H.~Fu,
  JHEP {\bf 1108}, 129 (2011)
  doi:10.1007/JHEP08(2011)129
  [arXiv:1105.3503 [hep-th]].
  
  
  
\bibitem{BjerrumBohr:2012mg} 
  N.~E.~J.~Bjerrum-Bohr, P.~H.~Damgaard, R.~Monteiro and D.~O'Connell,
  JHEP {\bf 1206}, 061 (2012)
  doi:10.1007/JHEP06(2012)061
  [arXiv:1203.0944 [hep-th]].
  
\bibitem{Fu:2012uy} 
  C.~H.~Fu, Y.~J.~Du and B.~Feng,
  JHEP {\bf 1303}, 050 (2013)
  doi:10.1007/JHEP03(2013)050
  [arXiv:1212.6168 [hep-th]].
  
\bibitem{Bjerrum-Bohr:2013iza} 
  N.~E.~J.~Bjerrum-Bohr, T.~Dennen, R.~Monteiro and D.~O'Connell,
  JHEP {\bf 1307}, 092 (2013)
  doi:10.1007/JHEP07(2013)092
  [arXiv:1303.2913 [hep-th]].
  
\bibitem{Du:2013sha} 
  Y.~J.~Du, B.~Feng and C.~H.~Fu,
  JHEP {\bf 1307}, 057 (2013)
  doi:10.1007/JHEP07(2013)057
  [arXiv:1304.2978 [hep-th]].
  
\bibitem{Tolotti:2013caa} 
  M.~Tolotti and S.~Weinzierl,
  JHEP {\bf 1307}, 111 (2013)
  doi:10.1007/JHEP07(2013)111
  [arXiv:1306.2975 [hep-th]].
  
\bibitem{Nohle:2013bfa} 
  J.~Nohle,
  Phys.\ Rev.\ D {\bf 90}, no. 2, 025020 (2014)
  doi:10.1103/PhysRevD.90.025020
  [arXiv:1309.7416 [hep-th]].
  
\bibitem{Monteiro:2013rya} 
  R.~Monteiro and D.~O'Connell,
  JHEP {\bf 1403}, 110 (2014)
  doi:10.1007/JHEP03(2014)110
  [arXiv:1311.1151 [hep-th]].
  
\bibitem{Fu:2014pya} 
  C.~H.~Fu, Y.~J.~Du and B.~Feng,
  JHEP {\bf 1408}, 098 (2014)
  doi:10.1007/JHEP08(2014)098
  [arXiv:1403.6262 [hep-th]].
  
\bibitem{Mastrolia:2015maa} 
  P.~Mastrolia, A.~Primo, U.~Schubert and W.~J.~Torres Bobadilla,
  Phys.\ Lett.\ B {\bf 753}, 242 (2016)
  doi:10.1016/j.physletb.2015.11.084
  [arXiv:1507.07532 [hep-ph]].
  
\bibitem{Fu:2016plh} 
  C.~H.~Fu and K.~Krasnov,
  JHEP {\bf 1701}, 075 (2017)
  doi:10.1007/JHEP01(2017)075
  [arXiv:1603.02033 [hep-th]].
  
\bibitem{Brown:2016mrh} 
  R.~W.~Brown and S.~G.~Naculich,
  JHEP {\bf 1610}, 130 (2016)
  doi:10.1007/JHEP10(2016)130
  [arXiv:1608.04387 [hep-th]].



\bibitem{Johansson:2014zca} 
  H.~Johansson and A.~Ochirov,
  JHEP {\bf 1511}, 046 (2015)
  doi:10.1007/JHEP11(2015)046
  [arXiv:1407.4772 [hep-th]].

\bibitem{Carrasco:2012ca} 
  J.~J.~M.~Carrasco, M.~Chiodaroli, M.~G\"unaydin and R.~Roiban,
  JHEP {\bf 1303}, 056 (2013)
  doi:10.1007/JHEP03(2013)056
  [arXiv:1212.1146 [hep-th]].

\bibitem{Chiodaroli:2013upa} 
  M.~Chiodaroli, Q.~Jin and R.~Roiban,
  JHEP {\bf 1401}, 152 (2014)
  doi:10.1007/JHEP01(2014)152
  [arXiv:1311.3600 [hep-th]].

\bibitem{Chiodaroli:2014xia} 
  M.~Chiodaroli, M.~G\"unaydin, H.~Johansson and R.~Roiban,
  JHEP {\bf 1501}, 081 (2015)
  doi:10.1007/JHEP01(2015)081
  [arXiv:1408.0764 [hep-th]].

\bibitem{Chiodaroli:2015rdg} 
  M.~Chiodaroli, M.~Gunaydin, H.~Johansson and R.~Roiban,
  JHEP {\bf 1706}, 064 (2017)
  doi:10.1007/JHEP06(2017)064
  [arXiv:1511.01740 [hep-th]].
  
\bibitem{Chiodaroli:2015wal} 
  M.~Chiodaroli, M.~Gunaydin, H.~Johansson and R.~Roiban,
  Phys.\ Rev.\ Lett.\  {\bf 117}, no. 1, 011603 (2016)
  doi:10.1103/PhysRevLett.117.011603
  [arXiv:1512.09130 [hep-th]].
  

  
\bibitem{Anastasiou:2016csv} 
  A.~Anastasiou, L.~Borsten, M.~J.~Duff, M.~J.~Hughes, A.~Marrani, S.~Nagy and M.~Zoccali,
  Phys.\ Rev.\ D {\bf 96}, no. 2, 026013 (2017)
  doi:10.1103/PhysRevD.96.026013
  [arXiv:1610.07192 [hep-th]].


\bibitem{Chiodaroli:2017ngp} 
  M.~Chiodaroli, M.~Gunaydin, H.~Johansson and R.~Roiban,
  JHEP {\bf 1707}, 002 (2017)
  doi:10.1007/JHEP07(2017)002
  [arXiv:1703.00421 [hep-th]].


\bibitem{Anastasiou:2017nsz} 
  A.~Anastasiou, L.~Borsten, M.~J.~Duff, A.~Marrani, S.~Nagy and M.~Zoccali,
  arXiv:1707.03234 [hep-th].
  

\bibitem{Chiodaroli:2017ehv} 
  M.~Chiodaroli, M.~Gunaydin, H.~Johansson and R.~Roiban,
  arXiv:1710.08796 [hep-th].
  
\bibitem{Bern:2011ia} 
  Z.~Bern and T.~Dennen,
  Phys.\ Rev.\ Lett.\  {\bf 107}, 081601 (2011)
  doi:10.1103/PhysRevLett.107.081601
  [arXiv:1103.0312 [hep-th]].

\bibitem{Bern:2011rj} 
  Z.~Bern, C.~Boucher-Veronneau and H.~Johansson,
  Phys.\ Rev.\ D {\bf 84}, 105035 (2011)
  doi:10.1103/PhysRevD.84.105035
  [arXiv:1107.1935 [hep-th]].

\bibitem{Bern:2012uf} 
  Z.~Bern, J.~J.~M.~Carrasco, L.~J.~Dixon, H.~Johansson and R.~Roiban,
  Phys.\ Rev.\ D {\bf 85}, 105014 (2012)
  doi:10.1103/PhysRevD.85.105014
  [arXiv:1201.5366 [hep-th]].

\bibitem{Bern:2012cd} 
  Z.~Bern, S.~Davies, T.~Dennen and Y.~t.~Huang,
  Phys.\ Rev.\ Lett.\  {\bf 108}, 201301 (2012)
  doi:10.1103/PhysRevLett.108.201301
  [arXiv:1202.3423 [hep-th]].
  
\bibitem{Bern:2012gh} 
  Z.~Bern, S.~Davies, T.~Dennen and Y.~t.~Huang,
  Phys.\ Rev.\ D {\bf 86}, 105014 (2012)
  doi:10.1103/PhysRevD.86.105014
  [arXiv:1209.2472 [hep-th]].
  

\bibitem{Bern:2013yya} 
  Z.~Bern, S.~Davies, T.~Dennen, Y.~t.~Huang and J.~Nohle,
  Phys.\ Rev.\ D {\bf 92}, no. 4, 045041 (2015)
  doi:10.1103/PhysRevD.92.045041
  [arXiv:1303.6605 [hep-th]].

\bibitem{Bern:2013qca} 
  Z.~Bern, S.~Davies and T.~Dennen,
  Phys.\ Rev.\ D {\bf 88}, 065007 (2013)
  doi:10.1103/PhysRevD.88.065007
  [arXiv:1305.4876 [hep-th]].
  
  
\bibitem{Bern:2013uka} 
  Z.~Bern, S.~Davies, T.~Dennen, A.~V.~Smirnov and V.~A.~Smirnov,
  Phys.\ Rev.\ Lett.\  {\bf 111}, no. 23, 231302 (2013)
  doi:10.1103/PhysRevLett.111.231302
  [arXiv:1309.2498 [hep-th]].
  

  
\bibitem{Bern:2014sna} 
  Z.~Bern, S.~Davies and T.~Dennen,
  Phys.\ Rev.\ D {\bf 90}, no. 10, 105011 (2014)
  doi:10.1103/PhysRevD.90.105011
  [arXiv:1409.3089 [hep-th]].

  
\bibitem{Bern:2014lha} 
  Z.~Bern, S.~Davies and T.~Dennen,
  arXiv:1412.2441 [hep-th].

\bibitem{Yang:2016ear} 
  G.~Yang,
  Phys.\ Rev.\ Lett.\  {\bf 117}, no. 27, 271602 (2016)
  doi:10.1103/PhysRevLett.117.271602
  [arXiv:1610.02394 [hep-th]].

\bibitem{Bern:2017puu} 
  Z.~Bern, H.~H.~Chi, L.~Dixon and A.~Edison,
  Phys.\ Rev.\ D {\bf 95}, no. 4, 046013 (2017)
  doi:10.1103/PhysRevD.95.046013
  [arXiv:1701.02422 [hep-th]].
  

  
\bibitem{Bern:2017tuc} 
  Z.~Bern, A.~Edison, D.~Kosower and J.~Parra-Martinez,
  Phys.\ Rev.\ D {\bf 96}, no. 6, 066004 (2017)
  doi:10.1103/PhysRevD.96.066004
  [arXiv:1706.01486 [hep-th]].
  
  
\bibitem{Johansson:2017bfl} 
  H.~Johansson, G.~K\"alin and G.~Mogull,
  JHEP {\bf 1709}, 019 (2017)
  doi:10.1007/JHEP09(2017)019
  [arXiv:1706.09381 [hep-th]].

  
\bibitem{Bern:2017ucb} 
  Z.~Bern, J.~J.~M.~Carrasco, W.~M.~Chen, H.~Johansson, R.~Roiban and M.~Zeng,
  arXiv:1708.06807 [hep-th].

\bibitem{Bern:2017yxu} 
  Z.~Bern, J.~J.~Carrasco, W.~M.~Chen, H.~Johansson and R.~Roiban,
  Phys.\ Rev.\ Lett.\  {\bf 118}, no. 18, 181602 (2017)
  doi:10.1103/PhysRevLett.118.181602
  [arXiv:1701.02519 [hep-th]].
  
\bibitem{Oxburgh:2012zr}
  S.~Oxburgh and C.~D.~White,
  JHEP {\bf 1302} (2013) 127
  doi:10.1007/JHEP02(2013)127
  [arXiv:1210.1110 [hep-th]].

\bibitem{Saotome:2012vy}
  R.~Saotome and R.~Akhoury,
  JHEP {\bf 1301} (2013) 123
  doi:10.1007/JHEP01(2013)123
  [arXiv:1210.8111 [hep-th]].

\bibitem{Vera:2012ds}
  A.~Sabio Vera, E.~Serna Campillo and M.~A.~Vazquez-Mozo,
  JHEP {\bf 1304} (2013) 086
  doi:10.1007/JHEP04(2013)086
  [arXiv:1212.5103 [hep-th]].
  
\bibitem{Melville:2013qca}
  S.~Melville, S.~G.~Naculich, H.~J.~Schnitzer and C.~D.~White,
  Phys.\ Rev.\ D {\bf 89} (2014) no.2,  025009
  doi:10.1103/PhysRevD.89.025009
  [arXiv:1306.6019 [hep-th]].

\bibitem{Johansson:2013nsa}
  H.~Johansson, A.~Sabio Vera, E.~Serna Campillo and M.~Á.~Vázquez-Mozo,
  JHEP {\bf 1310} (2013) 215
  doi:10.1007/JHEP10(2013)215
  [arXiv:1307.3106 [hep-th]].

\bibitem{Luna:2016idw}
  A.~Luna, S.~Melville, S.~G.~Naculich and C.~D.~White,
  JHEP {\bf 1701} (2017) 052
  doi:10.1007/JHEP01(2017)052
  [arXiv:1611.02172 [hep-th]].

\bibitem{Monteiro:2011pc} 
  R.~Monteiro and D.~O'Connell,
  JHEP {\bf 1107}, 007 (2011)
  doi:10.1007/JHEP07(2011)007
  [arXiv:1105.2565 [hep-th]].
  
\bibitem{Cheung:2016prv} 
  C.~Cheung and C.~H.~Shen,
  Phys.\ Rev.\ Lett.\  {\bf 118}, no. 12, 121601 (2017)
  doi:10.1103/PhysRevLett.118.121601
  [arXiv:1612.00868 [hep-th]].
  
\bibitem{Borsten:2013bp} 
  L.~Borsten, M.~J.~Duff, L.~J.~Hughes and S.~Nagy,
  Phys.\ Rev.\ Lett.\  {\bf 112}, no. 13, 131601 (2014)
  doi:10.1103/PhysRevLett.112.131601
  [arXiv:1301.4176 [hep-th]].
  
\bibitem{Anastasiou:2013hba} 
  A.~Anastasiou, L.~Borsten, M.~J.~Duff, L.~J.~Hughes and S.~Nagy,
  JHEP {\bf 1404}, 178 (2014)
  doi:10.1007/JHEP04(2014)178
  [arXiv:1312.6523 [hep-th]].

\bibitem{Anastasiou:2014qba} 
  A.~Anastasiou, L.~Borsten, M.~J.~Duff, L.~J.~Hughes and S.~Nagy,
  Phys.\ Rev.\ Lett.\  {\bf 113}, no. 23, 231606 (2014)
  doi:10.1103/PhysRevLett.113.231606
  [arXiv:1408.4434 [hep-th]].
  
\bibitem{Nagy:2014jza} 
  S.~Nagy,
  JHEP {\bf 1607}, 142 (2016)
  doi:10.1007/JHEP07(2016)142
  [arXiv:1412.4750 [hep-th]].
  
\bibitem{Anastasiou:2015vba} 
  A.~Anastasiou, L.~Borsten, M.~J.~Hughes and S.~Nagy,
  JHEP {\bf 1601}, 148 (2016)
  doi:10.1007/JHEP01(2016)148
  [arXiv:1502.05359 [hep-th]].
  
\bibitem{Cardoso:2016ngt} 
  G.~L.~Cardoso, S.~Nagy and S.~Nampuri,
  JHEP {\bf 1610}, 127 (2016)
  doi:10.1007/JHEP10(2016)127
  [arXiv:1609.05022 [hep-th]].
  

  
\bibitem{Cardoso:2016amd} 
  G.~Cardoso, S.~Nagy and S.~Nampuri,
  JHEP {\bf 1704}, 037 (2017)
  doi:10.1007/JHEP04(2017)037
  [arXiv:1611.04409 [hep-th]].
  
\bibitem{Holstein:2004dn}
  B.~R.~Holstein and J.~F.~Donoghue,
  Phys.\ Rev.\ Lett.\  {\bf 93} (2004) 201602
  doi:10.1103/PhysRevLett.93.201602
  [hep-th/0405239].

\bibitem{Ochirov:2014gsa} 
  A.~Ochirov,
  arXiv:1409.8087 [hep-th].

\bibitem{Bloch:1937pw}
  F.~Bloch and A.~Nordsieck,
  Phys.\ Rev.\  {\bf 52} (1937) 54.
  doi:10.1103/PhysRev.52.54

\bibitem{Kinoshita:1962ur}
  T.~Kinoshita,
  J.\ Math.\ Phys.\  {\bf 3} (1962) 650.
  doi:10.1063/1.1724268

\bibitem{Lee:1964is}
  T.~D.~Lee and M.~Nauenberg,
  Phys.\ Rev.\  {\bf 133} (1964) B1549.
  doi:10.1103/PhysRev.133.B1549



\bibitem{Laenen:2008gt}
  E.~Laenen, G.~Stavenga and C.~D.~White,
  JHEP {\bf 0903} (2009) 054
  doi:10.1088/1126-6708/2009/03/054
  [arXiv:0811.2067 [hep-ph]].

\bibitem{White:2011yy}
  C.~D.~White,
  JHEP {\bf 1105} (2011) 060
  doi:10.1007/JHEP05(2011)060
  [arXiv:1103.2981 [hep-th]].

\bibitem{Bern:1990cu}
  Z.~Bern and D.~A.~Kosower,
  Phys.\ Rev.\ Lett.\  {\bf 66} (1991) 1669.
  doi:10.1103/PhysRevLett.66.1669

\bibitem{Bern:1991aq}
  Z.~Bern and D.~A.~Kosower,
  Nucl.\ Phys.\ B {\bf 379} (1992) 451.
  doi:10.1016/0550-3213(92)90134-W

\bibitem{Strassler:1992zr}
  M.~J.~Strassler,
  Nucl.\ Phys.\ B {\bf 385} (1992) 145
  doi:10.1016/0550-3213(92)90098-V
  [hep-ph/9205205].

\bibitem{Vines:2017hyw} 
  J.~Vines,
  arXiv:1709.06016 [gr-qc].

\bibitem{Bern:1999ji} 
  Z.~Bern and A.~K.~Grant,
  Phys.\ Lett.\ B {\bf 457}, 23 (1999)
  doi:10.1016/S0370-2693(99)00524-9
  [hep-th/9904026].
  
\bibitem{Cheung:2016say} 
  C.~Cheung and G.~N.~Remmen,
  JHEP {\bf 1701}, 104 (2017)
  doi:10.1007/JHEP01(2017)104
  [arXiv:1612.03927 [hep-th]].
  
\bibitem{Cheung:2017kzx} 
  C.~Cheung and G.~N.~Remmen,
  JHEP {\bf 1709}, 002 (2017)
  doi:10.1007/JHEP09(2017)002
  [arXiv:1705.00626 [hep-th]].
  


\end{thebibliography}

\end{document}